%% file: fcbench.tex
\documentclass[sigconf, nonacm]{acmart}

\newcommand\vldbdoi{XX.XX/XXX.XX}
\newcommand\vldbpages{XXX-XXX}
\newcommand\vldbvolume{17}
\newcommand\vldbissue{7}
\newcommand\vldbyear{2024}
\newcommand\vldbauthors{\authors}
\newcommand\vldbtitle{\shorttitle} 
\newcommand\vldbavailabilityurl{https://github.com/hipdac-lab/FCBenchh}
\newcommand\vldbpagestyle{empty} 
\newcommand{\NA}{{-}}
\newcommand{\dm}[1]{\textcolor{black}{#1}} 
\usepackage{array}
\usepackage{tablefootnote}
\usepackage{graphicx}
\usepackage{xcolor}
\usepackage{enumitem}
\usepackage{amsmath}
\usepackage{tabu}
\usepackage{multirow}
\usepackage{caption}
\usepackage{subcaption}
\usepackage{balance}
\usepackage{xcolor,colortbl}
\usepackage[disable]{todonotes}
\usepackage{tcolorbox}
\usepackage{tikz}\usetikzlibrary{fpu}
\newcolumntype{^}{>{\currentrowstyle}}

\newcommand\NameDomain[2]{\makebox[2.5em][l]{#2}\hfill#1}

\newcommand{\TableHead}[1]{\multicolumn{1}{c}{\rotatebox{60}{#1}}}

\usepackage[hang]{footmisc}

\newcommand{\Software}[1]{\texttt{#1}}
\newcommand{\SoftwareAndRepo}[2]{\texttt{#1}\footnote{\texttt{#1}: \url{#2}}}

\newcommand{\TwoLineTabTitle}[2]{\begin{tabular}{@{}l@{}} #1\\[-1ex]#2 \end{tabular}}
\newcommand{\TabHead}{\bfseries\small}

\newcommand{\xc}[1]{\textcolor{black}{#1}}

\begin{document}
\title[FCBench]{FCBench: Cross-Domain Benchmarking of Lossless Compression for Floating-Point Data [Experiment, Analysis \& Benchmark]}

\settopmatter{authorsperrow=4}

\newcommand{\AFFIL}[4]{%
     \affiliation{%
         \institution{\small #1}
         \city{#2}\state{#3}\country{#4}
     }
     }

\author{Xinyu Chen}{\AFFIL{Washington State University}{Pullman}{WA}{USA}}
\email{xinyu.chen1@wsu.edu}

\author{Jiannan Tian}{\AFFIL{Indiana University}{Bloomington}{IN}{USA}}
\email{jti1@iu.edu}

\author{Ian Beaver}{\AFFIL{Verint Systems Inc}{Melville}{NY}{USA}}
\email{ian.beaver@verint.com}

\author{Cynthia Freeman}{\AFFIL{Verint Systems Inc}{Melville}{NY}{USA}}
\email{cynthiaw2004@gmail.com}

\author{Yan Yan}{\AFFIL{Washington State University}{Pullman}{WA}{USA}}
\email{yan.yan1@wsu.edu}

\author{Jianguo Wang}{\AFFIL{Purdue University}{West Lafayette}{IN}{USA}}
\email{csjgwang@purdue.edu}

\author{Dingwen Tao}{\AFFIL{Indiana University}{Bloomington}{IN}{USA}}
\authornote{Corresponding author: Dingwen Tao.}
\email{ditao@iu.edu}

\begin{abstract}
While both the database and high-performance computing (HPC) communities utilize lossless compression methods to minimize floating-point data size, a disconnect persists between them. Each community designs and assesses methods in a domain-specific manner, making it unclear if HPC compression techniques can benefit database applications or vice versa. With the HPC community increasingly leaning towards in-situ analysis and visualization, more floating-point data from scientific simulations are being stored in databases like Key-Value Stores and queried using in-memory retrieval paradigms. This trend underscores the urgent need for a collective study of these compression methods' strengths and limitations, \xc{not only} based on \xc{their performance in compressing} data from various domains \xc{but also on their runtime characteristics}. Our study extensively evaluates the performance of eight CPU-based and five GPU-based compression methods developed by both communities, using 33 real-world datasets assembled in the Floating-point Compressor Benchmark (FCBench). \xc{Additionally, we utilize the roofline model to profile their runtime bottlenecks.} Our goal is to offer insights into these compression methods that could assist researchers in selecting existing methods or developing new ones for integrated database and HPC applications.
\end{abstract}

\maketitle

\pagestyle{\vldbpagestyle}
\begingroup\small\noindent\raggedright\textbf{PVLDB Reference Format:}\\
\vldbauthors. \vldbtitle. PVLDB, \vldbvolume(\vldbissue): \vldbpages, \vldbyear.\\
\href{https://doi.org/\vldbdoi}{doi:\vldbdoi}
\endgroup
\begingroup
\renewcommand\thefootnote{}\footnote{\noindent
This work is licensed under the Creative Commons BY-NC-ND 4.0 International License. Visit \url{https://creativecommons.org/licenses/by-nc-nd/4.0/} to view a copy of this license. For any use beyond those covered by this license, obtain permission by emailing \href{mailto:info@vldb.org}{info@vldb.org}. Copyright is held by the owner/author(s). Publication rights licensed to the VLDB Endowment. \\
\raggedright Proceedings of the VLDB Endowment, Vol. \vldbvolume, No. \vldbissue\ %
ISSN 2150-8097. \\
\href{https://doi.org/\vldbdoi}{doi:\vldbdoi} \\
}\addtocounter{footnote}{-1}\endgroup

\ifdefempty{\vldbavailabilityurl}{}{
\vspace{.3cm}
\begingroup\small\noindent\raggedright\textbf{PVLDB Artifact Availability:}\\
The source code, data, and/or other artifacts have been made available at \url{\vldbavailabilityurl}.
\endgroup
}

\section{Introduction}
Floating-point data is widely used in various domains, such as scientific simulations, geospatial analysis, and medical imaging~\cite{habib2016hacc, saeedan2022spatial, fout2012adaptive}. As the scale of these applications increases, compressing floating-point data can help reduce data storage and communication overhead, thereby improving performance~\cite{roth1993database}.

\noindent\textbf{Why lossless compression?} Using a fixed number of bits (e.g., 32 bits for single-precision data) to represent real numbers often results in rounding errors in floating-point calculations~\cite{goldberg1991every}. Consequently, system designers favor using the highest available precision to minimize the problems caused by rounding errors~\cite{rubio2013precimonious}. Similarly, due to concerns about data precision, lossless compression is preferred over lossy compression, even with lower compression ratios, when information loss is not tolerable.

For instance, medical imaging data is almost always compressed losslessly for practical and legal reasons, while checkpointing for large-scale HPC simulations often employs lossless compression to avoid error propagation~\cite{fout2012adaptive}.
Lossless compression is also essential for inter-node communication in a majority of distributed applications~\cite{knorr2021ndzip}. This is because data is typically exchanged between nodes at least once per time step. Utilizing lossy compression would accumulate compression errors beyond acceptable levels, ultimately impacting the accuracy and correctness of the results.
Another example is astronomers often insist that they can only accept lossless compression because astronomical spectra images are known to be noisy~\cite{nieto1999data, du2009novel}. With the background (sky) occupying more than $95\%$ of the images, lossy compressions would incur unpredictable global distortions~\cite{oseret2007optimization}.

\subsection{Study Motivation}
Both the HPC and database communities have developed lossless compression methods for floating-point data. However, there are fundamental differences between the floating-point data of these two domains. Typically, numeric values stored in database systems do not necessarily \xc{display structural correlations except} for time-series data. \xc{In contrast}, HPC systems often deal with structured high-dimensional floating-point data produced by scientific simulations or observations, such as satellites and telescopes. The result is that the two communities have developed floating-point data compression methods for different types of data within their respective domains. Naturally, an intriguing question arises: \textit{Can the compression methods developed in one community be applied to the data from the other community, and vice versa?}

\xc{The urgency to answer this question arises from the growing trend of utilizing database tools on HPC systems. Practical use cases include in-situ visualization, which allows domain scientists to monitor and analyze large scientific simulations. For instance, \citeauthor{grosset2021lightweight} developed Seer-Dash, a tool that utilizes Mochi's Key-Value storage microservice \cite{ross2020mochi} and Google's LevelDB engine \cite{LevelDB} to generate in-situ visualizations for the HACC simulation \cite{habib2016hacc}.} \todo[color=blue!40]{R1\\W4}

To this end, we view the previously isolated evaluation of compression methods in each community on their data as a shortcoming. In this paper, we survey 13 CPU- and GPU-based lossless floating-point data compression software from both communities. \xc{We} evaluate their performances on 33 datasets from HPC, time-series, observation, and database transaction domains to fill the gap, \xc{while also profiling their runtime characteristics}. We aim to provide an efficient methodology for future HPC and database developers to select the most suitable compressor for their use case/application, thereby reducing the cost of trial and error.

\subsection{Our Contributions}
\begin{itemize}[noitemsep, topsep=2pt, leftmargin=1.3em]
    \item We present an experimental study of 13 lossless compression methods developed by the database and HPC communities for floating-point data.  
    \item We evaluate the performance of these 13 compression methods based on a wide variety of 33 real-world datasets, covering scientific simulations, time-series, observations, and database transaction domains. This helps to refresh our understanding of the selected methods, as they were previously evaluated only within their own data domains.
    \item We investigate the compression performance with different block/page sizes and measure the query overhead on a simulated in-memory database application. This provides insights for database developers to employ suitable compression methods.
    \item We utilize the roofline model~\cite{williams2009roofline} to assess the \xc{runtime characteristics} of selected algorithms regarding memory bandwidth utilization and computational operations, enabling potential to identify areas for performance enhancement.
    \item We \xc{employ statistical tools} to recommend the most suitable compression methods, considering various data domains, and considering both the end-to-end time and query overheads.
\end{itemize}


\noindent\textbf{Paper organization}.
The rest of this paper is organized as follows.
In \S\ref{sec:Background}, we present background about floating-point data, lossless compression techniques, and previous survey works. In \S\ref{sec:CPU} and \S\ref{sec:GPU}, we survey eight CPU-based and five GPU-based methods, respectively. In \S\ref{sec:expsetup} and \ref{sec:expresult}, we present the benchmarking methodology, experiment setup, and results. In \S\ref{sec:summary}, we summarize our findings and lessons.
In \S\ref{sec:conclude}, we conclude the paper and discuss future work.

\section{Background and Related Work}\label{sec:Background}
We introduce the background on floating-point data representation, lossless compression of floating-point data, and related studies.

\subsection{Floating-point Data}
\begin{figure}[h]
    \includegraphics[width=\linewidth,trim={0 0.45cm 0 0.2cm},clip]{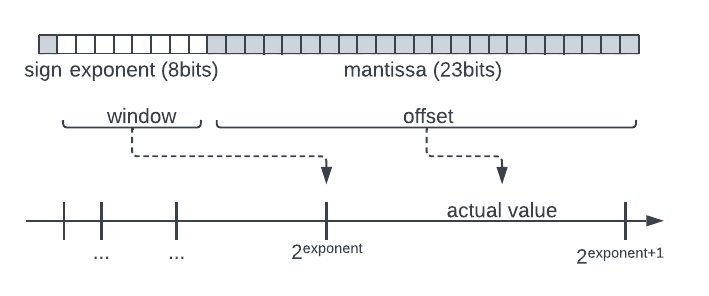}
    \vspace{-2mm}
    \caption{Single-precision format of the IEEE 754 standard.}
    \label{fig:ieee754}
    \vspace{-2mm}
\end{figure}
The IEEE 754 standard ~\cite{kahan1996ieee} defined floating-point data to be in 32-bit single word and 64-bit double word format. Figure~\ref{fig:ieee754} illustrates the single-precision format: one bit for the positive or negative \textit{sign}, 8 bits for the \textit{exponent}, and 23 bits for the \textit{mantissa}. The double-precision format includes one sign bit, 11 exponent bits, and 52 mantissa bits. The actual value of a floating-point datum is formulated as:
     $\text{x}=(-1)^\text{sign} \times 2^{(\text{exponent}-\text{bias})} \times 1.\text{mantissa}$. 

\subsection{Lossless Compression of Floating-point Data}
Lossless compression encodes the original data without losing any information. It is, therefore, used to compress text, medical imaging and enhanced satellite data~\cite{sayood2017introduction}, where information loss is not acceptable. Compression algorithms first identify the biased probability distribution, such as repeated patterns, and then encode the redundant information with reduced sizes. Some widely used encoding methods are listed below.
\begin{enumerate}[noitemsep, topsep=2pt, leftmargin=1.6em]
    \item Run-length coding replaces a string of adjacent equal values with the value itself and its count.
    \item Huffman coding builds optimal prefix codes to minimize the average length \cite{blelloch2001introduction}, based on the input data distribution.
    \item Arithmetic coding uses \textit{cumulative distribution functions} (CDF) to encode a sequence of symbols. It is more efficient than Huffman coding with increasing sequence length~\cite{sayood2017introduction}.
\end{enumerate}

\begin{figure}[h]
    \includegraphics[width=\linewidth,trim={2.0cm 6.5cm 0.5cm 7.0cm},clip]{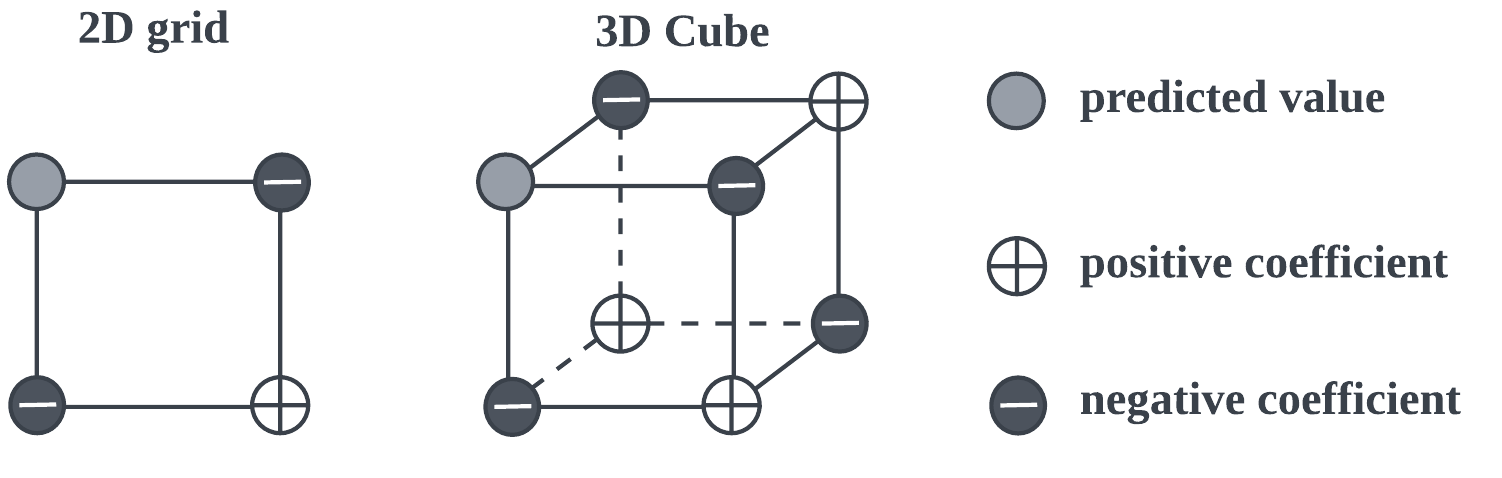}
    \vspace{-6mm}
    \caption{The Lorenzo transform and hypercubes.}
    \vspace{-2mm}
    \label{fig:lorenzo}
\end{figure}

\subsection{The Lorenzo Transform}
Data values from scientific simulations or sensors tend to correlate with neighboring values~\cite{knorr2021ndzip}. The Lorenzo predictor \cite{ibarria2003out} can leverage such structural dependencies to encode with fewer bits~\cite{cayoglu2019data}. Figure~\ref{fig:lorenzo} shows the value on a corner of a 2D grid or 3D cube can be estimated by the neighboring corners: $\hat{x}=\sum x_{\text{odd}} - \sum {x_{\text{even}}}$.
The 2D or 3D structures can be generalized to high-dimensional \textit{hypercubes} in the Lorenzo transform.
\vspace{-2mm}
\subsection{Friedman Test and Post-hoc Tests}
\xc{The machine learning community has long been embracing statistical validation to compare algorithms on a number of test data sets. According to the theoretical and empirical guidance by \citeauthor{demvsar2006statistical}, the Friedman test and corresponding post-hoc analysis overcome the limitations associated with using averaged metrics. These statistical tests are suitable to apply when the number of algorithms $k>5$ and the number of datasets $N>10$.} \todo[color=blue!40]{R4\\W5}
The Friedman test~\cite{demvsar2006statistical} compares the averaged ranks to find if all the algorithms are equivalent. \xc{The post-hoc Nemenyi test computes critical differences (CD) of averaged ranks. Finally, the CD diagram displays algorithms ordered by their average rank and groups of algorithms between which there is no significant difference.}

\subsection{Related Prior Surveys}
Previous surveys on data compression for databases and HPC applications have been conducted. ~\citeauthor{wang2017experimental} comprehensively studied 9 bitmap compression methods and 12 inverted list compression methods for databases on synthetic and real-world datasets~\cite{wang2017experimental}. However, bitmap and inverted list methods only compress categorical and integer values. \citeauthor{son2014data} studied 8 lossless and 4 lossy compression methods for scientific simulation checkpointing in the exascale era~\cite{son2014data}. They favored lossy compression algorithms for scientific simulations checkpointing but did not consider the situations when lossless compression is required. \citeauthor{lindstrom2017error} compared 7 lossless compression methods but only tested them on one turbulence simulation dataset~\cite{lindstrom2017error}. \xc{Although the evaluation metrics of space and time were common in previous studies from both communities, the database community distinguished itself by investigating the influence of compressors on query performances.}
\xc{Compared with previous benchmarks, our study stands out in three key aspects. \todo{Meta\\M2}\todo[color=blue!40]{R1\\W1} \textbf{First}, we include more recent compressors, covering multiple domains for a comprehensive evaluation of both software and datasets.  \textbf{Second}, we offer insights not only from the algorithm design perspective but also from the standpoint of system architecture.   \textbf{Third}, we employ a collection of tools to ensure a fair comparison of the selected methods. Specifically, we use statistical tests for rankings and recommendations, a simulated in-memory database for evaluating query performance, and the roofline model to investigate runtime bottlenecks.}
\begin{table*}[t]
    \caption{Summary of our studied lossless compression methods$^*$}
    \vspace{-2mm}
    \small
    \label{tab:compressor}
    \input{tables/selected_methods.tex}\\
    \footnotesize{\xc{$*$ bitshuffle methods (LZ4 and zstd) and nvcomp methods (LZ4 and bitcomp) are all listed separately.}}
    \footnotesize{\xc{$**$ S, D stand for single-/double-precision.}}
\end{table*}

\section{CPU-based compression methods} \label{sec:CPU}
\begin{figure}[h]
    \vspace{-2mm}
    \includegraphics[width=.92\linewidth,trim={00 0.5cm 0 0.5cm},clip]{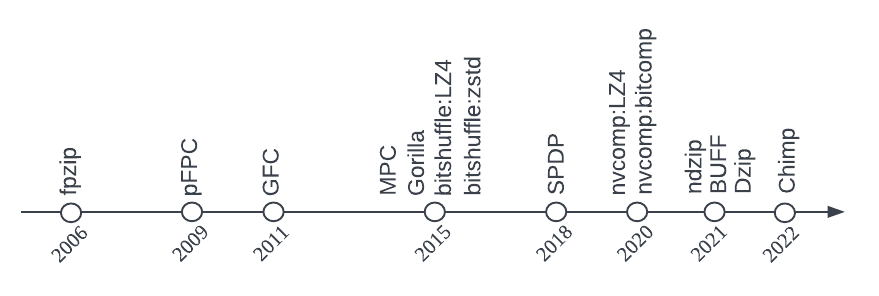}
    \caption{Timeline of studied compression methods.}
    \vspace{-4mm}
    \label{fig:survey_timeline}
\end{figure}

In this section, we describe eight CPU-based lossless compression methods. The first five are serial methods, and the last three are parallel methods. For each method, we introduce its computational workflow along with noteworthy features. Table~\ref{tab:compressor} summarizes all the studied methods. Their timeline is shown in Figure~\ref{fig:survey_timeline}.

\subsection{fpzip}
\noindent\textbf{fpzip~\cite{Lindstrom2006}} is a prediction-based compression algorithm that provides both lossless and lossy compression on the single- and double-precision floating-point data for scientific simulations.

\noindent\textbf{Workflow:} (1) fpzip uses the Lorenzo predictor~\cite{ibarria2003out} to predict the value of a hypercube corner from its previously encoded neighboring corners.  (2) The predicted and actual floating-point values are mapped to sign-magnitude integers to compute the residual. (3) The integer residuals' sign and leading zeros are symbols and encoded by a fast range coding method~\cite{martin1979range}. (4) The remaining non-zero bits are copied verbatim.

\noindent\textbf{Insights:} \todo[color=blue!40]{R1\\W3}For fpzip to achieve a better compression ratio, users should provide the Lorenzo predictor with the correct data dimensionality to predict with the hypercubes. Note that fpzip does not use parallel computing techniques.

\subsection{SPDP}\label{sec:spdp}
\noindent\textbf{SPDP~\cite{Claggett2018}} (Single Precision Double Precision) is a dictionary-based lossless compression algorithm for both precisions. It can work as an HDF5 filter~\cite{hdf5-filter} or a standalone compressor.

\noindent\textbf{Workflow:} SPDP is synthesized from three transform components to expose better data correlations and a reducer component to encode the transformed values. After sweeping over a total of 9,400,320 \footnote{The search space is $(\text{k}+1)(31+17)^{\text{k}-1}\times17 = 9400320$ for combining  $\text{k}=4$ components from 31 transform candidates and 17 reducer candidates.} combinations on 26 scientific datasets, the authors selected the following four components that rendered the best compression ratios: (1) \texttt{LNVs2} subtracts the last $2^{nd}$ byte value from the current byte value and emits the residual. (2) \texttt{DIM8} groups the most significant bytes of the residuals, followed by the second most significant bytes, etc. This puts the exponent bits into consecutive bytes. (3) \texttt{LNVs1} computes the difference between the previously grouped consecutive bytes. (4) \texttt{LZa6} is a fast variant of the LZ77~\cite{ziv1977universal} to encode the final residuals.

\noindent\textbf{Insights:} SPDP has the trade-off between compression ratio and throughput because \texttt{LZa6} employs a sliding window to encode the positions and lengths of matched patterns.  Larger sliding window sizes can increase the compression ratio with the cost of decreased throughput due to prolonged searching time.


\subsection{BUFF}
\noindent\textbf{BUFF~\cite{Liu2021}} is a delta-based compression algorithm for low-precision floating-point data, commonly used in server monitoring and IoT (Internet of Things) devices. Two features distinguish BUFF from other methods in this survey. (1) Without precision information, BUFF essentially becomes a lossy compressor.  (2) BUFF can directly query byte-oriented columnar encoded data without decoding. This capability allows BUFF to achieve a speedup ranging from 35x to 50x for selective and aggregation filtering.

\begin{table}[]
    \caption{Number of bits needed for targeted precision in BUFF.}
    \vspace{-2mm}
    \label{tab:buffbits}
    \input{tables/buffbits}
    \vspace{-2mm}
\end{table}

\noindent\textbf{Workflow:} (1) BUFF splits the input values into integer and fractional components. (2) It then uses a look-up table (Table~\ref{tab:buffbits}) to keep the most significant bits of the mantissa part and discard the trailing bits. (3) BUFF computes the difference between the current and minimal values. (5) BUFF uses padding to encode the integer and fraction parts into a multiple of bytes. Each byte unit is treated as a sub-column and stored together to enable query operations. (6) The value range and precision information are saved as metadata and compressed data for decompression.

\noindent\textbf{Insights:} BUFF's compression ratio is sensitive to the value ranges and outliers. It does not employ parallel processing. The byte-column query follows the pattern match method. Assume each data point $x$ is encoded and saved in sub-columns $x_1, x_2, ..., x_k$. To perform a predicate $x == C$, BUFF translates $C$ into sub-columns $C_1, C_2, ..., C_k$ and evaluates the equal operator on the sub-columns one at a time. BUFF will skip a record once a sub-column is disqualified ($x_i != C_i$). 
\subsection{Gorilla}
\noindent\textbf{Gorilla~\cite{Pelkonen2015}} is a delta-based lossless compression algorithm to compress the timestamps and the data values for the in-memory time series database used at Facebook. 

\noindent\textbf{Workflow:} Given that time series data are often represented as pairs of a timestamp and a value, Gorilla uses two different methods: (1) It uses delta-of-delta to compress timestamps.  With the fixed interval of time series data, the majority of timestamps can be encoded as a single bit of \texttt{0}. (2) For floating-point data values, Gorilla conducts XOR operations on the current and previous values and encodes the residuals from the XOR operation. (3) Gorilla uses a single bit $C=\texttt{0}$ to store all-zero residuals; $C=\texttt{10}$ is used to store the actual \textit{meaningful bits} when the nonzero bits of the residual fall within the block bounded by the previous leading zeros and trailing zeros; $C=\texttt{11}$ uses $5$ bits for the length of leading-zeros, $6$ bits for the length of meaningful bits and then the actual residual.

\noindent\textbf{Insights:} Gorilla's performance is sensitive to the data patterns. The overhead of control bits becomes high when data values change frequently.  Gorilla does not employ parallel computing techniques.
\subsection{Chimp}
\noindent\textbf{Chimp~\cite{liakos2022chimp}} is a lossless compression algorithm to compress floating-point values of time series data. Based on Gorilla's~\cite{Pelkonen2015} workflow, Chimp redesigned the control bits to improve compression ratio when the trailing zeros of XORed residuals are less than $6$. Furthermore, Chimp computes the best XORed residual (having the highest number of trailing zeros) from the $128$ previous values. In other words, Chimp is a prediction-based method with a sliding window (\cite{Burtscher2007, Burtscher2009}).

\noindent\textbf{Workflow:} (1) Chimp maintains evicting queues to store 128 previous values grouped by their less significant bits. This enables Chimp to get more trailing zeros from the XORed residuals than Gorilla. (2) Chimp uses control bits $C=\texttt{00}$ to encode all-zero residuals; For $C=\texttt{01}$, Chimp uses 3 bits to encode the length of leading zeros and 6 bits to encode the length of meaningful bits. Then it stores the meaningful bits; For $C=\texttt{10}$, the current length of leading zeros equals the previous length of leading zeros, so Chimp directly stores the meaningful bits; For $C=\texttt{11}$, Chimp uses 3 bits to encode the length of leading zeros, then stores the meaningful bits.

\noindent\textbf{Insights:} Using a sliding window allows Chimp to achieve a higher compression ratio when data values are more random. However, the overhead of looking up the sliding window also decreases Chimp's compression throughput compared with Gorilla's.

\subsection{pFPC}
\noindent\textbf{pFPC~\cite{Burtscher2009}} is a prediction-based algorithm that losslessly compresses double-precision scientific simulation data in parallel.

\noindent\textbf{Workflow:} (1) pFPC stores historical value sequences in two hash tables and predicts current values by looking up the hash tables. (2) The residuals are computed from the XOR operation between the actual and predicted values. (3) The leading zeros of the XORed result and the selected hash predictor are encoded with 4 bits. (4) The nonzero residual bytes are copied. 

\noindent\textbf{Insights:} pFPC utilizes parallel computing to increase the overall throughput. The original data is partitioned into chunks and distributed across multiple CPU threads (\xc{default 8 pthreads})\todo[color=blue!40]{R4\\W1}. However, there exists a trade-off between compression ratio and throughput. Given that high-dimensional scientific data often exhibits higher correlation along the same dimension, pFPC prefers to align the number of threads with the data dimensionality. With a large number of threads and big chunk sizes, mixing values from multiple dimensions can decrease the compression ratio. Thus, pFPC requires data dimensionality as input parameters and typically does not fully utilize multi-threading capacity to align with the data dimensionality.

\subsection{Bitshuffle} \label{sec:bitshuffle}
\textbf{Bitshuffle}~\cite{Masui2015} is a data transform in itself. The algorithm can expose the data correlations within a subset of bits in a byte to improve the compression ratio for downstream encoders.

\noindent\textbf{Workflow:} (1) Bitshuffle splits the input data into blocks and distributes the blocks among threads. On each thread, a block's \textit{bits} are arranged into a $m \times n$ matrix, where $m$ is the number of values in a block, and $n$ is the data element size (32 or 64 bits). Then, it performs a bit-level transpose to get a $n \times m$ matrix, where the i$^{th}$ ($1 \le i \le n$) bits are combined into bytes. (2) Bitshuffle uses other methods, such as LZ4 and zstd, to encode the transposed data. 

\noindent\textbf{Insights:} Bitshuffle employs SSE2 and AVX2 vectorized instruction sets for parallelized transforms. Although larger blocks will increase the compression ratios, the default block size is set to $4096$ bytes to ensure that Bitshuffle can fit data into the L1 cache. LZ4 and zstd are then applied to the cached data to improve performance.
\subsection{ndzip-CPU} \label{sec:ndzip-c}
\noindent\textbf{ndzip~\cite{Knorr2021}} is a prediction-based lossless compression algorithm. The CPU implementation uses SIMD instructions and thread-level parallelism to achieve high throughput.

\noindent\textbf{Workflow:} (1) ndzip divides data into blocks, with each block corresponding to a hypercube containing 4096 elements. (2) A multidimensional Integer Lorenzo transform computes the residuals within each block. (3) The residuals are divided into chunks of $32$ single-precision or $64$ double-precision values and performs bit-transpose operations. (4) The zero-words in the transposed chunks are removed. The positions of zero-words are encoded with 32- or 64-bit bitmap headers, and the non-zero words are copied. 

\noindent\textbf{Insights:} ndzip employs multi-level parallel computing to achieve high throughputs. The hypercubes are independently compressed using thread-level parallelism, while each hypercube's transformation, transposition, and encoding leverage SIMD vector instructions.  
\section{GPU-based Compression Methods} \label{sec:GPU}
\subsection{GFC} \label{sec:GFC}
\noindent\textbf{GFC~\cite{o2011floating}} is a delta-based compression algorithm for double-precision floating-point scientific data. GFC leverages the massive GPU parallelism to achieve high compression/decompression throughput.

\noindent\textbf{Workflow:} (1) GFC divides input data into chunks equal to the number of GPU warps, each consisting of 32 threads. These chunks are further divided into subchunks of $32$ double-precision values and compressed independently. (2) GFC computes residuals by subtracting the last value of the previous subchunk from the current subchunk. (3) GFC encodes the sign and leading zeros with 4 bits followed by the non-zero residual bytes. These operations are performed in parallel across GPU warps.

\noindent\textbf{Insights:}  GFC sets the size of subchunks to $32$ to align with the number of GPU threads in a warp.  However, the delta-based predictor sacrifices accuracy to accommodate multidimensional data within fixed-sized (256 Bytes) subchunks. This is due to the computation of all residuals for the current 32 values by subtracting the last value from the previous 32. Another limitation of GFC is that the input data size cannot exceed $512$ MB based on the hardware available during their research.
\subsection{MPC}
\noindent\textbf{MPC~\cite{yang2015mpc}}, which stands for Massive Parallel Compression, is a synthesized delta-based lossless compression algorithm for floating-point data. It is constructed from four components following the process described in \S\ref{sec:spdp}. The number of combinatorial search spaces for MPC is $138,240$.

\noindent\textbf{Workflow:} MPC divides input data into chunks of 1024 elements and processes them in parallel. The pipeline consists of four components: (1) \texttt{LNV6s} computes the residual by subtracting the $6^{th}$ prior value in the same chunk from the current value. (2) \texttt{BIT} reorganizes the data chunks by emitting the most significant bit of each word first (packing them into words), followed by the second most significant bits, and so on. This is essentially the same operation of Bitshuffle~\cite{Masui2015}.  (3) \texttt{LNV1s} computes differences between consecutive words from the BIT transform. Finally, (4) \texttt{ZE} outputs a bitmap to indicate zero values in the chunk and copies the non-zero values after the bitmap.

\noindent\textbf{Insights:} MPC resembles ndzip in the entire pipeline, except for using the delta-based predictor to replace the Lorenzo prediction. The input word size (single- or double-precision) information is important so that \texttt{LNV6s} computes the correct residuals. MPC demonstrated that a delta-based predictor could achieve good compression ratios when combined with data transform components.
\subsection{nvCOMP}
\noindent\textbf{nvCOMP}~\cite{githubnvcomp} is a CUDA library by NVIDIA to provide APIs of compressors and decompressors. In the latest version (2.4), nvCOMP includes $8$ compression algorithms. We include \texttt{nvCOMP::LZ4} because it achieves the highest compression ratio among the 8 compressors. Similarly, we include \texttt{nvCOMP::bitcomp} for it has the highest compression/decompression throughput.

\noindent\textbf{Workflow:} nvCOMP has been proprietary software since version 2.3, and NVIDIA does not describe the detailed workflow for each compression method.

\noindent\textbf{Insights:} nvCOMP::LZ4 and \texttt{nvCOMP::bitcomp} do not require extra input parameters such as dimensional information.
\subsection{ndzip-GPU}
{ndzip-GPU~\cite{knorr2021ndzip}} is the GPU-based parallelization scheme for the ndzip algorithm in \S\ref{sec:ndzip-c}. While the algorithm remains the same, the GPU implementation further improves parallelism by distributing transforming and residual coding among up to 768 threads.

\noindent\textbf{Workflow:} The compression pipeline of ndzip-GPU is the same as the CPU implementation: (1) Divide data into hypercubes, (2) Compute residuals using the Lorenzo predictor, (3) Perform bit-transposition, (4) Remove zero-words, output the bit-map header and uncompressed non-zeros. 

\noindent\textbf{Insights:} ndzip-GPU first writes encoded chunks to a global scratch to guarantee the order of variable-length encoded chunks. After computing a parallel prefix sum to obtain the offsets for all chunks, ndzip copies the encoded chunks from scratch memory to the output stream. By retrieving the offsets for each compressed block, the decompression is fully block-wise parallel without synchronization.
\subsection{Dzip}
While Dzip~\cite{Goyal2021} is not developed by the HPC or database community, the Neural Network-based compression method represents an emerging direction. Therefore, we include this method to enrich our survey.  Dzip is a general-purpose lossless compressor in contrast to the aforementioned floating-point oriented methods. It trains two recurrent neural network (RNN) models to estimate the conditional distribution for input data symbols and encodes them with an arithmetic encoding method.

\noindent\textbf{Workflow:} (1) An RNN-based bootstrap model is trained for multiple passes. (2) Dzip trains a larger supporter model and the bootstrap model in a single pass to predict the conditional probability of the current symbol. The supporter models are discarded after encoding to save disk space. (3) Dzip retrains a new supporter model combined with the bootstrap model in one pass during decoding.  This supporter model is again discarded later.

\noindent\textbf{Insights:} The bootstrap model is trained and saved. However, the supporter model needs retraining for unseen datasets. Although Dzip is faster than other NN-based compressors, such as NNCP~\cite{bellard2021nncp} and CMIX~\cite{cmixgithub}, its compression speed is about several KB/s. Thus, NN-based compression methods are still not practical for applications at the time of our survey.

\section{Benchmark Methodology and Setup}\label{sec:expsetup}
In this section, we present our benchmarking methodology and experimental setups, including test datasets, software and hardware configurations, and evaluation metrics.
\subsection{Our Methdology}
\label{sec:methodology}
This benchmarking aims to enrich our understanding of these compression algorithms' performance from various perspectives. Our benchmarking study encompasses three aspects. 
\subsubsection{The compression aspect:} We evaluate the selected algorithms with generic metrics: compression ratio, (de)compression throughputs, end-to-end wall time, the effect of dimensionality parameters, and scalability of parallel compression. For fair comparisons, we perform statistical tests on the rankings. 

\subsubsection{The database query aspect:}\label{sec:micro-benchmarking} 
\begin{figure}[t]
    \centering
    \includegraphics[
        ,width=\linewidth
        ,trim={3.5in 2.1in 5in 3in}
        ,clip
    ]{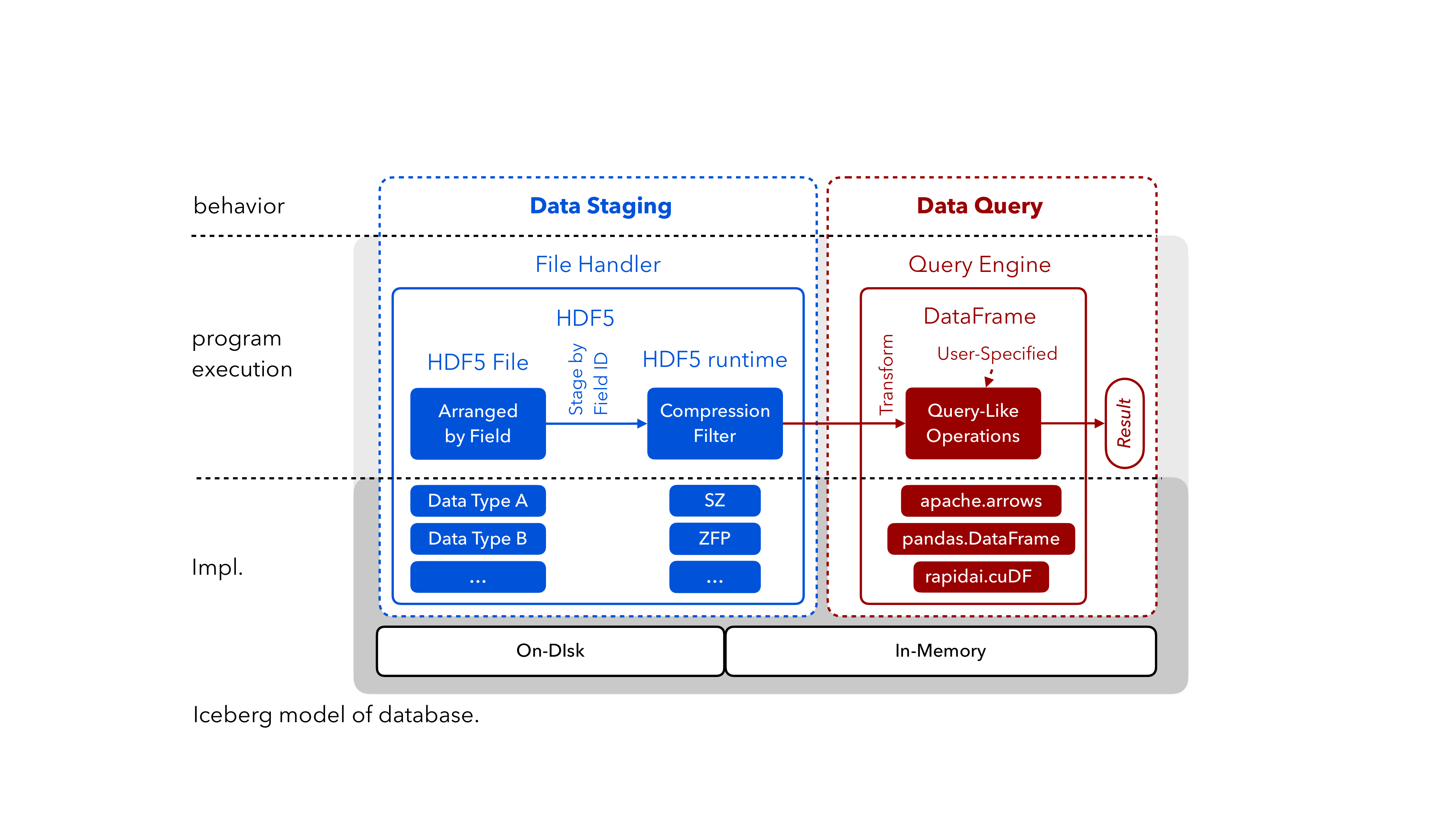}
    \vspace{-4mm}
    \caption{Integrating HPC and database with HDF5 and Dataframes.}
    \vspace{-2mm}
    \label{fig:integrated-workflow}
\end{figure}

\xc{We adopt a micro-benchmarking approach \cite{lissandrini2018beyond, boral1984methodology} to swiftly evaluate query performance with compression. Rather than analyzing queries in a comprehensive database system, we develop a tool to simulate an in-memory database and investigate three primitive operations: file I/O, data decoding, and full table scan query}. \todo[color=blue!40]{R1\\W2}
More concretely, we depict an HPC system in Figure \ref{fig:integrated-workflow} that reads Hierarchical Data Format 5 (HDF5) \cite{folk2011overview} files from the disk into Pandas \cite{snider2004pandas} dataframes in memory and performs the queries on these in-memory dataframes.  This simulated database enables us to quickly check the compression performance with various block sizes and measure the decompression overhead for query operations under the database context.  

\xc{Our simulated tool has limitations\todo[color=blue!40]{R1\\W2} in accurately reflecting the true query performance of integrated compression methods in a real database system. This is mainly because a full table scan oversimplifies the process compared to more complex operations like join and update queries in in-situ visualization applications.} it does help to bypass the substantial engineering efforts needed to integrate compressors into an actual database system, aiding in the selection of the best-fit method.

\subsubsection{The system design aspect:} We use the roofline model~\cite{williams2009roofline} to identify potential bottlenecks, such as arithmetic intensity and memory bandwidth, for future compression algorithm developers.

\subsection{Evaluation Metrics}
We use the compression ratio (CR), compression throughput (CT), and decompression throughput (DT) to measure the compression performance. They are calculated as follows.
\begin{flalign*}
    \text{CR}  =\frac{\text{orig size}}{\text{comp size}}, \quad
    \text{CT}  =\frac{\text{orig size}}{\text{comp time}}, \quad
    \text{DT}  =\frac{\text{orig size}}{\text{decomp time}}
\end{flalign*}

For fpzip, pFPC, Gorilla, SPDP, ndzip (CPU and GPU), BUFF, Chimp, GFC, and MPC, we measured the times by adding instructions before and after the compression/decompression function to exclude the I/O. For bitshuffle (LZ4 and zstd) and nvcomp (LZ4 and bitcomp), we directly reported the timings and compression ratio obtained from their built-in benchmark functions. We repeated each method on the selected datasets ten times and reported each dataset's average compression ratios, run times, and throughputs. We used the harmonic mean of compression ratios and arithmetic mean of throughputs to evaluate the overall performance.d.

\subsection{Datasets}
We choose 33 open datasets from four domains, shown in Table~\ref{tab:data}. The datasets include (1) scientific simulation data from Scientific Data Reduction Benchmarks \cite{zhao2020sdrbench} and \cite{knorr2020datasets};
(2) time series data such as sensor streams, stock market, and traffic data which typically require fewer precision digits; (3) observation data such as HDR photos and telescope images; (4) simulated data generated from the Transaction Processing Performance Council Benchmark (TPC)~\cite{council2005transaction}, including numeric columns extracted from the TPC-H, TPCX-BB, and TPC-DS transactions. \xc{ Although the largest data size of 4GB in our selection is much small for a typical HPC application, it is a common workload of a time step for in-situ analysis \cite{grosset2021lightweight} and has been evaluated by many benchmark studies \cite{zhao2020sdrbench, knorr2020datasets}.\todo[color=blue!40]{R4\\W4} }  

\begin{table}
    \caption{Evaluated floating-point datasets.}
    \label{tab:data}
    \resizebox{\linewidth}{!}{%
        \input{tables/datasets_new}
    }
    \footnotesize{\xc{$**$ S, D stand for single-/double-precision.}}
\end{table}

\subsection{Friedman Test and Post-hoc Tests}
Following recent survey works \cite{freeman2021experimental, holder2023review}, we apply the Friedman test and apply the CD diagram \cite{demvsar2006statistical} to compare the selected compression methods. \footnote{\xc{The number of algorithms and datasets are big enough.}} We use $\alpha=0.05$, $k=13$, $N=33$ for the hypothesis test and compute the critical difference. \todo[color=blue!40]{R4\\W5}

\subsection{System Configuration}
We evaluate the selected compression methods on a Chameleon Cloud \cite{keahey2020lessons} compute node with two Intel Xeon Gold 6126 CPUs (2.6 GHz), 187 GB RAM and one Nvidia Quadro RTX 6000 GPU (24 GB VRAM). The node is configured with Ubuntu 20.04, GCC/G++ 9.4, CUDA 11.3, CMAKE 3.25.0, HDF5 1.14.1, and Python 3.8.

\SoftwareAndRepo{fpzip}{https://github.com/LLNL/fpzip},
\SoftwareAndRepo{pFPC}{https://userweb.cs.txstate.edu/~burtscher/research/pFPC/},
SPDP, 
\SoftwareAndRepo{ndzip-CPU}{https://github.com/celerity/ndzip} are compiled with GCC/G++ $9.4$;
\Software{\texttt{nvCOMP::LZ4}} and \SoftwareAndRepo{\texttt{nvCOMP::bitcomp}}{https://developer.nvidia.com/nvcomp-download} are executable binary files downloaded from the benchmark page; \Software{bitshuffle::LZ4} and \SoftwareAndRepo{bitshuffle+zstd}{https://github.com/kiyo-masui/bitshuffle.git} are compiled with Python 3.8 and GCC 9.4; Gorilla and Chimp are integrated in \SoftwareAndRepo{influxdb}{https://github.com/panagiotisl/influxdb}, where they can be compiled with go 1.18.0 and rustc 1.53.0;  \SoftwareAndRepo{BUFF}{https://github.com/Tranway1/buff} is compiled with the nightly version of rust.
\SoftwareAndRepo{GFC}{https://userweb.cs.txstate.edu/~burtscher/research/GFC/},
\SoftwareAndRepo{MPC}{https://userweb.cs.txstate.edu/~burtscher/research/MPC/} and
\SoftwareAndRepo{ndzip-GPU}{https://github.com/celerity/ndzip} are compiled with nvcc 11.3.  Each C/C++ method was compiled with \texttt{-O3} flag. When possible, we changed the compile flags for GPU-based methods to use the current GPU architecture \texttt{-arch=sm\_75}.

\newtcbox{\xmybox}[1][red]{on line,
arc=3pt,colback=#1!10!white,colframe=#1!50!black,
before upper={\rule[-3pt]{0pt}{10pt}},boxrule=1pt,
boxsep=0pt,left=1pt,right=2pt,top=0pt,bottom=0pt}

\section{Evaluation Results}\label{sec:expresult}
In this section, we present the results following our evaluation methodology described in \S\ref{sec:methodology}. We begin by discussing general compression performance, covering compression ratio, compression throughput, decompression throughput, and end-to-end wall time (including memory copy, especially for GPU-based compressors). This is followed by detailed discussions. Subsequently, we delve into the evaluation results related to compression performance in the context of databases. Finally, we present the findings from the roofline model.

\subsection{General Compression Performance}
We evaluate the compression algorithms using the following parameters: (1) the compression level, which is set for the best compression ratios (CRs), and (2) blocks/chunks, which are set to default sizes when applicable.
\subsubsection{Compression Ratio}\label{sec:exp_cr}
\begin{figure}[ht] 
    \includegraphics[width=.9\linewidth]{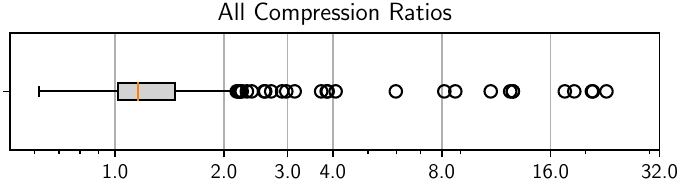}
    \vspace{-4mm}
    \caption{BoxPlot of compression ratios.}
    \vspace{-2mm}
    \label{fig:cr_overview}
\end{figure}

\hfill\\
\xmybox[gray]{\textbf{Observation 1: compression ratios $\le 2.0$}}  Figure~\ref{fig:cr_overview} shows all measured compression ratios. The median is $1.16$ and outliers range from $2.0$ to $22.8$. Our experiments support previous study ~\cite{lindstrom2017error} that floating-point data is difficult to compress. 

\begin{figure}[ht] 
    \begin{subfigure}{\linewidth}
        \includegraphics[width=.95\linewidth]{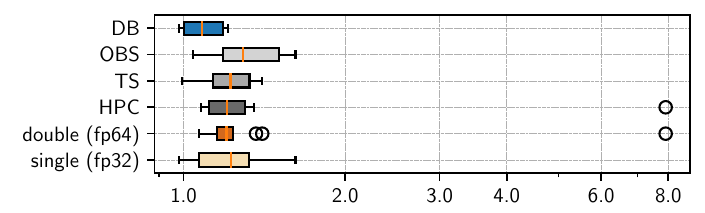}
        \vspace{-4mm}
        \caption{Data type and domains}
        \label{fig:cr_data}
    \end{subfigure}
    \hfill
    \begin{subfigure}{\linewidth}
        \includegraphics[width=.95\linewidth]{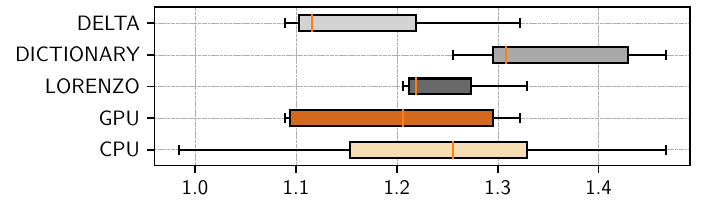}
        \caption{Prediction methods and hardware platforms}
        \label{fig:cr_method}
        \vspace{-2mm}
    \end{subfigure}
    \caption{Compression ratios by data/methods groups.}
    \vspace{-2mm}
\end{figure}

Figure~\ref{fig:cr_data} shows the compression ratios on datasets of different domains and data types. (1) The median compression ratio is 1.225 for single- and 1.202 for double-precision datasets. This result supports the study \cite{yang2015mpc} that double-precision data are more challenging to compress. (2) Observation datasets (OBS) have the highest median compression ratio of $1.292$, followed by HPC, Time Series (TS), and Database (DB) datasets with those of $1.206$, $1.223$, and $1.080$. DB is the most difficult domain to compress.

Figure~\ref{fig:cr_method} shows the compression ratios of different groups of compression methods based on their predictors and hardware platforms. (1) Dictionary-based methods (bitshuffle::LZ4, bitshuffle+zstd, Chimp) achieve higher compression ratios than Lorenzo-based methods (fpzip, ndzip-CPU, ndzip-GPU) and Delta-based methods (Gorilla, GFC, and MPC). The median compression ratios for these three groups are 1.309, 1.219, and 1.116, respectively. (2) On the selected datasets, CPU-based methods exhibit higher compression ratios compared to GPU-based methods.

\xc{\textbf{Analysis:} \todo{Meta\\M1}\todo[color=blue!40]{R1\\W3}(1) Except for BUFF, the selected algorithms compress either leading zeros or zero words. This is intuitive, as the exponents are more compressible. Double-precision data are less compressible due to their larger mantissa portion. (2) The OBS dataset achieves the best compression ratio as it consists of highly structured single-precision values. In contrast, the DB dataset is the most challenging to compress due to its lack of structural patterns. (3) Dictionary-based predictors outperform others because they search for patterns over a longer range. (4) CPU-based methods tend to use more dictionary-based predictors compared to GPU-based methods.}

\xc{\textbf{Takeaway:} (1) Using single-precision for saving numeric values in databases is desirable. (2) Dictionary-based predictors perform better than delta- and Lorenzo-based predictors.} \\
\begin{figure}
    \begin{subfigure}{\linewidth}
        \centering
        \includegraphics[width=.9\linewidth]{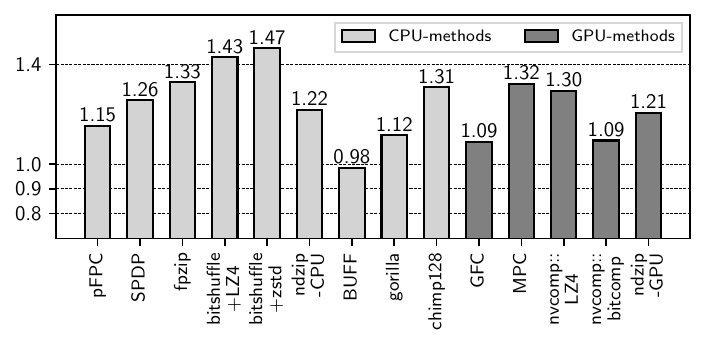}
        \vspace{-2mm}
        \caption{Average compression ratios}
        \label{fig:cr_per_method}
    \end{subfigure}
    \hfill
    \begin{subfigure}{\linewidth}
        \centering
        \includegraphics[width=.95\linewidth,trim={0 0 0 0},clip]{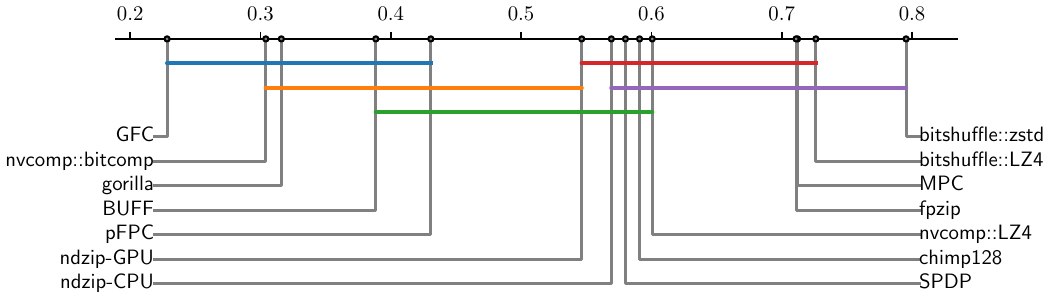}
        \caption{critical difference diagram on compression ratios}
        \label{fig:cr_cddiagram}
        \vspace{-2mm}
    \end{subfigure}
    \caption{\xc{Compression ratios of different methods.}}
    \vspace{-4mm}
\end{figure} 
\xmybox[gray]{\textbf{Observation 2: No significant winner}} \todo[color=blue!40]{R4\\W5}Table~\ref{tab:comp_ratio} shows the detailed CRs and Figure~\ref{fig:cr_per_method} shows their harmonic mean CRs.  \xc{In Figure~\ref{fig:cr_cddiagram}, we present the critical difference diagram, where the selected methods are ordered according to their average rankings (higher ranking is better) and cliques (a group of algorithms) are connected by a line representing the critical difference. We further observe: (1) The CD diagram shows no clear winner because bitshuffle+zstd is not significantly better than SPDP, although it has a significantly higher CR than ndzip-GPU.} (2) GFC ranks the lowest but is not significantly lower than nvCOMP::bitcomp, gorilla, BUFF, and pFPC. (3) fpzip has the highest compression ratio on HPC datasets; nvCOMP::LZ4 works the best on Time series (TS); bitshuffle+zstd performs the best for Observation (OBS) datasets; Chimp performs the best on Database (DB) datasets. (4) CPU-based methods are more robust than GPU-based methods: $2.0\%$ of CPU experiments incurred runtime errors, while $7.3\%$ of the GPU experiments were killed by runtime errors.

\xc{\textbf{Analysis:} (1) Dictionary-based predictors not only help bitshuffle::zstd achieve the top rank in compression ratios (CRs), but they also assist Chimp128 in outperforming Gorilla. (2) Bit-level transpose operations can expose subtle patterns, such as identical i$^{th}$ bits, benefiting both bitshuffle and MPC. (3) As described in \S\ref{sec:GFC}, the less accurate predictor of GFC contributes to its lower ranking.}

\xc{\textbf{Takeaway:} (1) Although the critical difference does not significantly distinguish bitshuffle::zstd from the rest in the group (including bitshuffle::Lz4, MPC, fpzip, nvCOMP::LZ4, Chimp128, and SPDP), it does highlight a direction for future research: exploring bit-level transposition and dictionary-based predictors to improve compression ratios. (2) For highly structured data from the HPC or OBS domains, delta- and Lorenzo-based compressors are comparable to, or even better than, dictionary-based compressors. (3) As numerical values in DB datasets lack structure, dictionary-based methods outperform delta- and Lorenzo-based methods.}
\begin{table*}
    \caption{Compression ratios (i.e., $\text{original size}/\text{compressed size}$).}
    \vspace{-2mm}
    \label{tab:comp_ratio}
    \resizebox{1.0\linewidth}{!}{%
        \input{tables/comp_ratio_new}%
    }
    \vspace{-2mm}
\end{table*}

\subsubsection{Compression Throughput}\label{sec:CT}
\begin{figure}[t]
    \begin{subfigure}{\linewidth}
        \includegraphics[width=.95\linewidth]{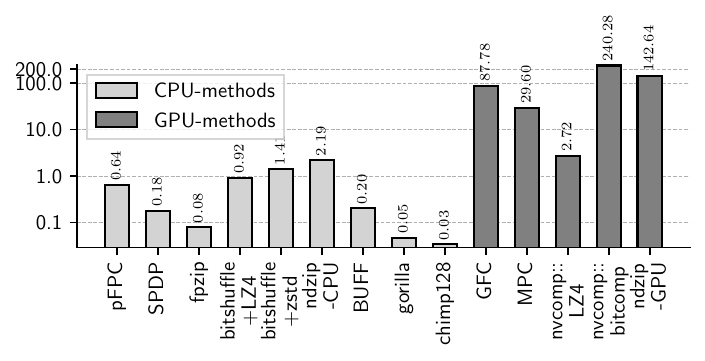}
        \vspace{-4mm}
        \caption{Compression throughputs.}
        \label{fig:cthr_per_method}
    \end{subfigure}
    
    \begin{subfigure}{\linewidth}
        \includegraphics[width=.95\linewidth]{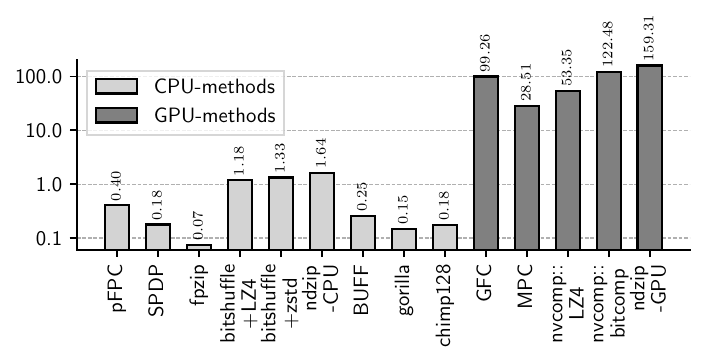}
        \vspace{-4mm}
        \caption{Decompression throughputs.}
        \vspace{-2mm}
        \label{fig:dthr_per_method}
    \end{subfigure}
    \caption{(De)compression throughputs of different methods.}
    \vspace{-4mm}
\end{figure}

\hfill\\
\xmybox[gray]{\textbf{Observation 3: GPU-based method is 350x faster}} The median of compression throughput for GPU- and CPU-based methods are 73.71 GB/s and 0.21 GB/s respectively. 
Figure \ref{fig:cthr_per_method} and Table~\ref{tab:comp_speed} show the average compression throughputs of selected methods. We further observe that (1) Among our selection, nvCOMP::bitcomp and ndzip-CPU are the fastest GPU- and CPU-based method respectively.  (2) nvCOMP::LZ4 is the slowest GPU-based methods. (3) Although serial methods such as Chimp, Gorilla, and fpzip are significantly slower, BUFF has a decent compression speed.

\xc{\textbf{Analysis:} (1) The LZ4 algorithm can cause significant branch divergence on GPUs, slowing down the compression process. Consequently, nvCOMP::LZ4 exhibits the lowest compression ratio (CR) compared to other GPU-based methods that utilize delta or Lorenzo predictors. (2) bitshuffle::zstd, bitshuffle::LZ4, and ndzip-CPU all leverage SIMD instructions and thread-level parallelism to enhance compression throughputs. This advantage helps them surpass pFPC, which relies solely on pthreads for parallel computation. (3) Among serial methods, BUFF stands out for its speed, attributable to the efficiency of the Rust language \cite{rust-lang, bugden2022rust}.}

\xc{\textbf{Takeaway:}  Dictionary-based methods are more prone to branch divergences, a significant challenge for GPU architectures. In addition, the high performance of the Rust language is noteworthy.} 

\begin{table*}
    \caption{Compression \& decompression throughput (GB/s).}
    \vspace{-2mm}
    \label{tab:comp_speed}
    \small\sffamily
    \resizebox{\linewidth}{!}{%
        \input{tables/all_avg_throughputs}%
    }
    \vspace{-2mm}
\end{table*}

\begin{table*}
    \caption{End-to-end wall time (ms).}
    \small
    \vspace{-2mm}
    \label{tab:endtoend}
    \sffamily
    \input{tables/end_to_end}
    \\[1ex]
    $*$We omit two nvcomp methods because the benchmark binary has no API to measure standalone walltime without I/O.
    \vspace{-2mm}
\end{table*}


\subsubsection{Decompression Throughput}\label{sec:DT}
\begin{figure}[t]
    \includegraphics[width=.95\linewidth]{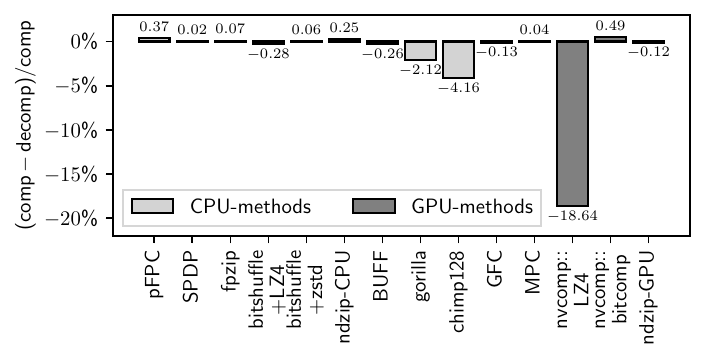}
    \vspace{-6mm}
    \caption{$r_D = \frac{CT-DT}{CT}$. Positive $r_D$ means compression is faster.}
    \vspace{-2mm}
    \label{fig:diff_thr}
\end{figure}

\hfill\\
\xmybox[gray]{\textbf{Observation 4: Decompression is faster}}
Not surprisingly, Figure~\ref{fig:dthr_per_method} shows that GPU-based methods have higher decompression throughput. A bit more interesting observation is shown in Figure~\ref{fig:diff_thr} that decompression is faster than compression on average. We further observe that (1) ndzip-GPU and ndzip-CPU are the fastest GPU- and CPU-based method respectively, while fpzip has the lowest DT ($0.07$ GB/s). (2) Dictionary-based methods have higher decompression speed than their compression. For nvCOMP::LZ4 and Chimp128, DT are 18.64$\times$ and 4$\times$ of their CT. However, bitshuffle+zstd and bitshuffle::LZ4 have balanced CT and DT. (3) Delta and Lorenzo based methods have balanced compression and decompression speed.

\xc{\textbf{Analysis:} (1) Dictionary-based methods require significantly fewer computations during decoding, leading to LZ4 algorithm and Chim128 exhibiting higher decompression times (DT) than compression times (CT). (2) In contrast, delta and Lorenzo predictors perform more balanced operations for both compression and decompression. (3) bitshuffle::zstd and bitshuffle::LZ4 demonstrate balanced CT/DT, behaving more similarly to delta and Lorenzo-based methods. We will later show that they are memory-bound using the roofline model.}

\xc{\textbf{Takeaway:} Dictionary-based methods do not cause many branch divergences during decompression. Their faster decompression speed is advantageous for query operations, as databases often decompress data multiple times.}

\subsubsection{End-to-End Wall Time}\label{sec:walltime}\hfill\\
\xmybox[gray]{\textbf{Observation 5: Host-to-device is slow}}
Table~\ref{tab:endtoend} shows the wall time of the selected methods and includes the overhead of memory copy from hosts to GPU cards. In terms of end-to-end time, bitshuffle::LZ4 and bitshuffle::zstd are comparable with GFC and MPC, while ndzip-CPU is even faster than ndzip-GPU.

\xc{\textbf{Takeaway:} The overhead of host-to-device memory copy is nonnegligible. } \xc{Bitshuffle::zstd has the best overall computation time plus its highest averaged CR in \S\ref{sec:exp_cr}. }

\subsubsection{Effect of Dimension Information}\hfill\\
Delta and Lorenzo-based methods require dimension information to improve prediction accuracy. In column-based databases, high-dimensional data are stored as 1-d columns. We try to compress the multi-dimensional datasets as 1-d arrays and employ Mann-Whitney U Test \cite{nachar2008mann} ($\alpha=0.05$) to test significant change of CRs. \\
\xmybox[gray]{\textbf{Observation 6: Compression is 1-d friendly}}
Table~\ref{tab:cr_md1d} compares the CRs with/without the dimension information.  The Mann-Whitney U Test finds no significant difference.

\xc{\textbf{Analysis:} (1) Treating high-dimensional data as 1-D arrays causes the Lorenzo predictor to degrade to the delta predictor, which is capable of exposing data correlations. With the aid of bit-level transpose operations, the compression ratios (CRs) of MPC, ndzip-CPU, and ndzip-GPU do not exhibit significant changes. (2) The GFC predictor remains inaccurate, even with the correct dimension information because the residuals are computed from the current chunk and the last value in the previous chunk.}

\xc{\textbf{Takeaway:} Column-based databases can effectively utilize delta- and Lorenzo-based compression methods as they scale the prediction errors uniformly. Additionally, a bit-level transpose operation can further reduce the impact of these prediction errors.}

\newcommand{\Entry}[4][black]{%
\begin{tikzpicture}
\draw[fill=#1, fill opacity=.2, draw=#1!50] ({-log2(1+#3*0.2)},-0.08) rectangle (0, -0.02);
\node[anchor=south east, inner sep=0pt, outer sep=0pt] at (0, 0) {
\footnotesize
    \begin{tabular}{@{}r@{}} 
        \normalsize #2 MB/s \\[-1ex]
        \footnotesize #3$\times$ (#4)
    \end{tabular}
};
\end{tikzpicture}%
}%

\begin{table}[htbp]
\renewcommand{\arraystretch}{1.2}
  \centering\small\sffamily
  \label{tab:compscale}
  \resizebox{0.8\linewidth}{!}{%
  \begin{tabular}{ r | r|r|r|r| }
\multicolumn{1}{r}{thread \#} &	
\multicolumn{1}{c}{pFPC}	&	
\multicolumn{1}{c}{Bitshfl+LZ4}	&	
\multicolumn{1}{c}{Bitshfl+Zstd}	&	
\multicolumn{1}{c}{ndzip-CPU}	
\\
\cline{2-5}
1	&	\Entry{133}{1.00}{100\%}	&	\Entry{997}{1.00}{100\%}	&	\Entry{250}{1.00}{100\%}	&	\Entry{1655}{1.00}{100\%}	\\
2	&	\Entry{172}{1.29}{65\%}	&	\Entry{1562}{1.57}{78\%}	&	\Entry{470}{1.88}{94\%}	&	\Entry{1640}{0.99}{50\%}	\\
4	&	\Entry{225}{1.69}{42\%}	&	\Entry{2420}{2.43}{61\%}	&	\Entry{869}{3.48}{87\%}	&	\Entry{1658}{1.00}{25\%}	\\
8	&	\Entry{352}{2.65}{33\%}	&	\Entry{3413}{3.42}{43\%}	&	\Entry{1545}{6.18}{77\%}	&	\Entry{1682}{1.02}{13\%}	\\
16	&	\Entry{530}{3.98}{25\%}	&	\Entry{3547}{3.56}{22\%}	&	\Entry{2432}{9.73}{61\%}	&	\Entry{1682}{1.02}{6\%}	\\
24	&	\Entry{618}{4.65}{19\%}	&	\Entry{2977}{2.98}{12\%}	&	\Entry{2739}{10.96}{46\%}	&	\Entry{1683}{1.02}{4\%}	\\
32	&	\Entry{523}{3.93}{12\%}	&	\Entry{2699}{2.71}{8\%}	&	\Entry{2098}{8.40}{26\%}	&	\Entry{1678}{1.01}{3\%}	\\
48	&	\Entry{531}{3.99}{8\%}	&	\Entry{1583}{1.59}{3\%}	&	\Entry{1551}{6.21}{13\%}	&	\Entry{1678}{1.01}{2\%}	\\
\cline{2-5}
\end{tabular}
}
\caption{\xc{Parallel compression throughputs.}}

\resizebox{0.8\linewidth}{!}{%
  \label{tab:decompscale}
  \begin{tabular}{ r | r|r|r|r| }
\multicolumn{1}{r}{thread \#} &	
\multicolumn{1}{c}{pFPC}	&	
\multicolumn{1}{c}{Bitshfl+LZ4}	&	
\multicolumn{1}{c}{Bitshfl+Zstd}	&	
\multicolumn{1}{c}{ndzip-CPU}	
\\
\cline{2-5}
1	&	\Entry{91}{1.00}{100\%}	&	\Entry{1746}{1.00}{100\%}	&	\Entry{1135}{1.00}{100\%}	&	\Entry{1197}{1.00}{100\%}	\\
2	&	\Entry{110}{1.20}{60\%}	&	\Entry{2797}{1.60}{80\%}	&	\Entry{1889}{1.67}{83\%}	&	\Entry{1205}{1.01}{50\%}	\\
4	&	\Entry{151}{1.65}{41\%}	&	\Entry{4021}{2.30}{58\%}	&	\Entry{2938}{2.59}{65\%}	&	\Entry{1198}{1.00}{25\%}	\\
8	&	\Entry{224}{2.46}{31\%}	&	\Entry{4874}{2.79}{35\%}	&	\Entry{4068}{3.59}{45\%}	&	\Entry{1223}{1.02}{13\%}	\\
16	&	\Entry{306}{3.36}{21\%}	&	\Entry{3463}{1.98}{12\%}	&	\Entry{3527}{3.11}{19\%}	&	\Entry{1208}{1.01}{6\%}	\\
24	&	\Entry{367}{4.02}{17\%}	&	\Entry{2698}{1.55}{6\%}	&	\Entry{2704}{2.38}{10\%}	&	\Entry{1210}{1.01}{4\%}	\\
32	&	\Entry{367}{4.02}{13\%}	&	\Entry{2450}{1.40}{4\%}	&	\Entry{2461}{2.17}{7\%}	&	\Entry{1220}{1.02}{3\%}	\\
48	&	\Entry{353}{3.88}{8\%}	&	\Entry{1365}{0.78}{2\%}	&	\Entry{1356}{1.19}{2\%}	&	\Entry{1217}{1.02}{2\%}	\\
\cline{2-5}
\end{tabular}
}

  \caption{\xc{Parallel decompression throughputs.}}
  \vspace{-6mm}
\end{table}%

\subsubsection{\xc{Scalability of Parallel Compression}}\todo[color=blue!40]{R4\\W2}\hfill\\
\xmybox[gray]{\textbf{Observation 7: Parallel compressors can scale up}}
\xc{We measure the scalability of the compressors that support parallel/multi-thread mode, noting that a data parallel design can effectively scale up with multiple threads. Table-\ref{tab:decompscale} shows that they can achieve  3$\sim$4$\times$ speedup with 16 to 24 threads compared to their single-thread performances. However, ndzip-CPU does not exhibit similar scalability, which may be due to its implementation issue.}

\begin{table*}
    \caption{Dimension information's influence on compression ratios.}
    \vspace{-2mm}
    \label{tab:cr_md1d}
    \small\sffamily
        \input{tables/comp_ratio_md1d}
    \vspace{-1mm}
\end{table*}


\begin{figure}[b]
    \centering
    \vspace{-4mm}
    \includegraphics[width=.95\linewidth]{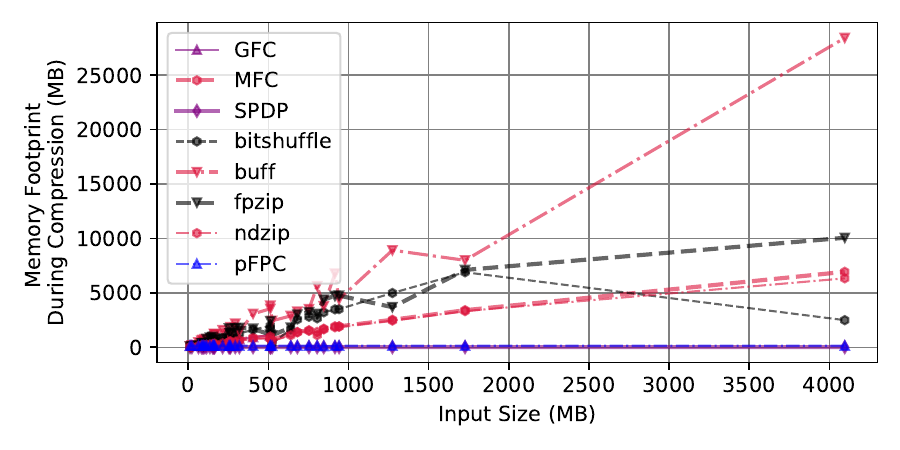}
    \vspace{-6mm}
    \caption{\xc{Memory footprints with different input data sizes.}}
    \label{fig:memfootp}
\end{figure}

\subsubsection{\xc{Memory Footprint}}\todo[color=blue!40]{R4\\W2}\hfill\\
\xc{As shown in Figure \ref{fig:memfootp}, while most of the selected compressors use a memory footprint approximately twice the size of the input data, pFPC and SPDP have fixed sizes for their read and write buffers, resulting in a constant memory footprint across all datasets. In contrast, BUFF incurs a memory footprint that is 7$\times$ larger than the input data, rendering it less suitable for in-situ analysis.}    

\subsection{Performance under Context of Databases}
Following the micro-benchmarking approach in \S\ref{sec:micro-benchmarking}, we investigate the performance of selected compression methods with different block/chunk sizes in a simulated database system. The compressed data are initially stored in HDF5 files. They are read from the disk and decompressed into Pandas dataframes. Finally, we perform a table scan query. 

\subsubsection{Performance under Different Block Sizes}\hfill\\
\begin{table*}
    \caption{Compression performance under different block sizes.}
    \vspace{-2mm}
    \label{tab:blocksizes}
    \small
    \sffamily
    \resizebox{0.70\linewidth}{!}{%
        \input{tables/blocksizes}
    }
\end{table*}

HDF5 datasets consist of multiple disk pages (chunks) \cite{sqlitepagesize} similar to database page files. The default page size ranges from $4$KB to $8$KB. Note that compression algorithms utilizes larger block sizes from $64$KB to $8$MB.

\xmybox[gray]{\textbf{Observation 8: Compressors prefer larger block sizes}} Table~\ref{tab:blocksizes} displays the average CR, CT, DT of 8 compression algorithms \footnote{We omit algorithms that cannot be easily converted to work with blocks.} with $4$KB, $64$KB, and $8$MB as block sizes. The best combinations of (method, metric) are highlighted in bold. Seven out of eight compression algorithms yield improved CRs, and all algorithms exhibit higher throughts with $64$KB- and $8$MB- block sizes.

\xc{\textbf{Takeaway:} We suggest database designers to increase the default page sizes to improve the compression performances.}

\subsubsection{Query Performance on TPC Datasets}\hfill\\
\begin{table*}
    \caption{Read and query time (in ms) from HDF5 files.}
    \vspace{-2mm}
    \label{tab:query_time}
    \sffamily
    \resizebox{0.96\linewidth}{!}{%
        \input{tables/query-from-h5}
    }
\end{table*}

To investigate query performance, we measure the running time of three primitive operations depicted in Figure~\ref{fig:integrated-workflow}: (1) file
I/O time to retrieved compressed data from HDF5 files \cite{folk2011overview}; (2) data decoding time;  (3) full table scan query on the Pandas dataframes \cite{snider2004pandas}. Table~\ref{tab:query_time} presents the average reading and query time for selected compression algorithms on the TPC benchmark datasets.   For each TPC dataset, the reading overhead varies due to different CRs and DTs based on the compression algorithms; while the query time remains consistent, as the retrieved Pandas dataframes are identical across all algorithms. For instance, pFPC spends $78$ ms to read the compressed chunks and $356$ ms to decompress the TPCH-order data. Subsequently, Pandas uses $190$ ms to query the dataframe\footnote{The average query time is measured on a set of full table scans: \texttt{df.loc[df.A<=$v_i$]}, where $v_i$ are from the histogram of \texttt{df.A}. The number of histogram bins is 10.}.

\xmybox[gray]{\textbf{Observation 9: End-to-end time is important}} The query performance aligns with the end-to-end wall time of each method. Despite of its fast query speed, we do not recommend GFC because of its limitation on input data sizes.

\xc{\textbf{Takeaway:} We highlight bitshuffle+zstd as the prime CPU-based compressor and MPC as the foremost GPU-based compressor. }

\subsection{Performance Analysis via Roofline Model}
In this section, we examine the the runtime bottlenecks from the standpoint of compression developers. \footnote{We utilize Intel Advisor \cite{IntelAdvisor} and Nsight Compute \cite{NvidiaNsight} to profile on the msg-bt data.} The roofline model \cite{williams2009roofline} visualizes algorithms as dots under the roof, which is formed by the peak memory bandwidth and computations.

\begin{figure*}[ht]
    \begin{subfigure}{.49\linewidth}
        \includegraphics[width=.95\textwidth]{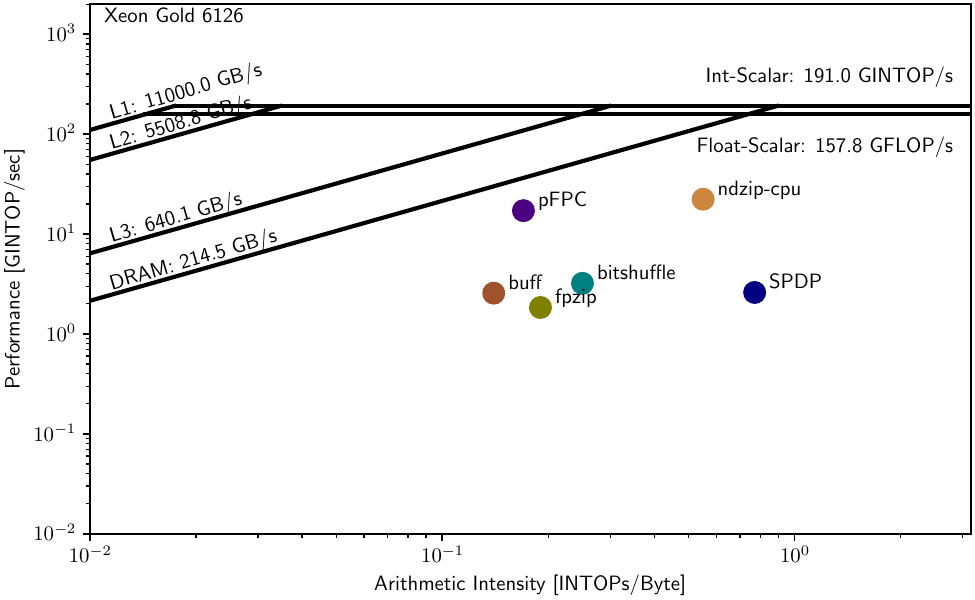}
        \caption{CPU-based methods}
        \vspace{-2mm}
        \label{fig:rf_cpu}
    \end{subfigure}
    \hfill
    \begin{subfigure}{.49\linewidth}
        \includegraphics[width=.95\textwidth]{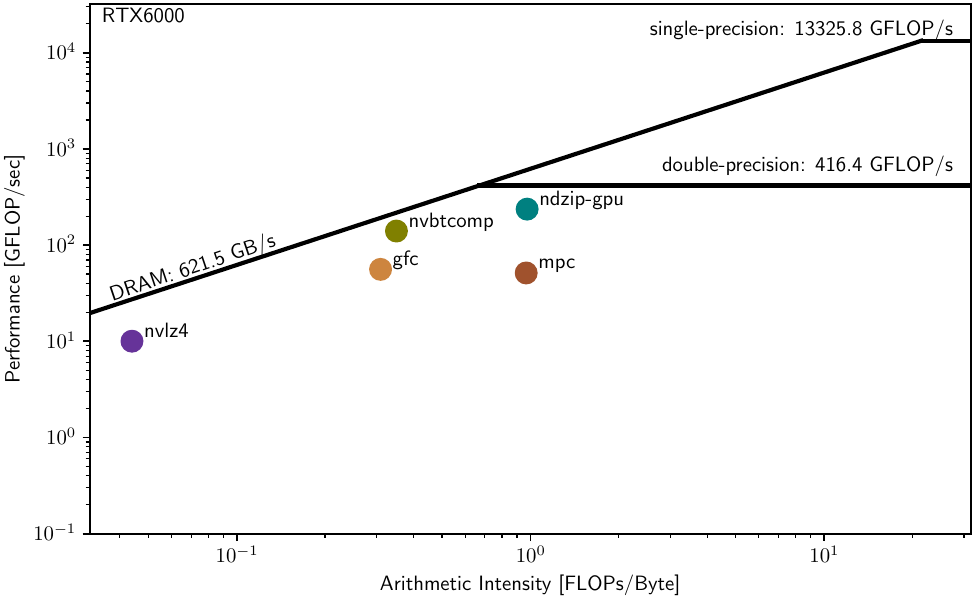}
        \caption{GPU-based methods}
        \vspace{-2mm}
        \label{fig:rf_gpu}
    \end{subfigure}
    \caption{Roofline analysis for different methods.}
    \vspace{-4mm}
\end{figure*}

\xmybox[gray]{\textbf{Observation 10: Three potential improvements}}  Figure \ref{fig:rf_cpu} and \ref{fig:rf_gpu} shows the performance of CPU- and GPU-based methods respectively\footnote{We omit Gorilla and Chimp for go is not supported in Advisor.}.  Each dot represents the most expensive function/loop that consumes greater than $40\%$ computation time of the corresponding methods. We further observe that (1) Most GPU-based methods are close to the memory bandwidth roof.   (2) Some CPU-based methods are far below the roof.

\xc{\textbf{Analysis:} (1) Serial methods (BUFF, fpzip, SPDP) are not bound by memory or computation. The introduction of parallel techniques could potentially improve their throughputs. (2) The throughput of bitshuffle::LZ4 and bitshuffle::zstd could be enhanced by increasing the number of threads, as evidenced by the scalability test. (3) ndzip::CPU and ndzip::GPU are computation bound. Potential improvements should consider reducing branch divergence.}

\xc{\textbf{Takeaway:} The roofline model is an effective tool for providing performance insights into the selected algorithms. For instance, it suggests that the throughputs of bitshuffle methods can be improved by utilizing more threads, circumventing the need for costly scalability tests.}

\vspace{-1mm}
\section{Summary}\label{sec:summary}\todo[color=blue!40]{R3\\W4}
In this section, we share the valuable insights from our study. We reflect on key lessons learned, highlight essential takeaways, and conclude with specific recommendations based on our evaluations.

\vspace{-1mm}
\subsection{Lessons Learned}
\textbf{First}, understanding the characteristics of data is crucial as it allows us to use single-precision values to achieve higher compression ratios when feasible. \textbf{Second}, while GPU-based methods offer faster computation, the overheads associated with host-to-device data movement and branch divergence often become the bottlenecks.

\vspace{-2mm}
\subsection{Key Takeaways}\todo{Meta\\M1}\todo[color=blue!40]{R3\\W1}
\xc{\textbf{For compressor developers:} \todo[color=blue!40]{R1\\W1}\todo[color=blue!40]{R3\\W3}The trade-off between ratio and throughput remains a crucial consideration. When data exhibit a structural layout, such as in scientific simulations and observational data, delta and Lorenzo-based predictors are effective in compression ratios and can easily employ data parallel designs. Conversely, when data display repeated patterns, such as in time series data and database transactions, dictionary-based methods are more effective, albeit at the cost of challenging GPU parallelism due to branch divergence. Ultimately, data transforms like bit and byte-level shuffling effectively improve compression ratios.}

\xc{\textbf{For database designers:} Compression methods show promise in reducing disk storage, as many algorithms offer fast decompression speeds and can compress 1-D arrays for column-based databases without degrading compression ratio performance. To fully utilize these compression algorithms, it is advisable to set larger default database page sizes.}

\xc{\textbf{For system architects:} Using the roofline model can reveal that increasing the number of threads and employing parallel computations can improve compression/decompression speeds without the need for expensive scalability tests.} 

\vspace{-3mm}
\subsection{Recommendations}
\xc{\textbf{For users focused on storage reduction:} based on the rankings in \S\ref{sec:exp_cr}, we recommend fpzip for HPC data, nvCOMP::LZ4 for time series data, bitshuffle::zstd for observational data, and Chimp for database data. \textbf{For users needing fast speed:} We suggest bitshuffle::LZ4, bitshuffle::zstd, MPC, and ndzip-CPU/GPU, as they demonstrate short end-to-end times in \S\ref{sec:walltime}. \textbf{For general users:} We recommend bitshuffle::zstd and MPC due to their balanced performance in average compression ratio (CR), swift end-to-end wall time, and minimal retrieval overhead for queries. Overall, bitshuffle methods rank as the top choice due to their better robustness and lower cost of CPU hardware.}
    .

\vspace{-1mm}
\section{Conclusion}\label{sec:conclude}
In this study, we conducted a comprehensive examination of 13 lossless floating-point data compressors across 33 datasets from multiple domains. Our analysis delves into their performance, considering not only algorithmic aspects but also architectural considerations. By employing a combination of statistical testing, the roofline model, and a simulated in-memory database, we scrutinized the key designs of the selected compressors. Building on this analysis, we offer recommendations for both compressor researchers and database architecture designers. Additionally, we have created a map to assist users in selecting the most suitable compressors based on their specific requirements. These efforts represent our commitment to bridging the gap between independently developed compression methods in the HPC and database communities.  



%
\begin{acks}
This work is partly supported by NSF Awards \#2247080, \#2303064 \#2311876, and \#2312673.
Results presented in this paper were obtained using the Chameleon testbed supported by NSF.
\end{acks}

\newpage

\bibliographystyle{ACM-Reference-Format}
\bibliography{fcbench}

\end{document}

%% file: tables/selected_methods.tex
\sffamily
\begin{tabu}{@{} r|c|c| >{\ttfamily}c|c|c|c|c|c|c|c|c| @{}}
 \rowfont{\bfseries}
\multicolumn{1}{c}{}	& \multicolumn{1}{c}{year} & \multicolumn{1}{c}{domain} &\multicolumn{1}{c}{\sffamily precision$^{**}$} & \multicolumn{1}{c}{arch.} & \multicolumn{1}{c}{parallel impl.} & \multicolumn{1}{c}{language} & \multicolumn{1}{c}{trait}      & \multicolumn{1}{c}{availability} \\
	\cline{2-9}
	{fpzip \cite{Lindstrom2006}}          & 2006           & HPC                       & {S,D}               & CPU             & serial                   & C++                & Lorenzo              & open-source
	\\
	{pFPC \cite{Burtscher2009}}           & 2009           & HPC                       & {\ \ D}             & CPU             & threads                  & C                  & prediction           & open-source
	\\
	{bitshuffle::LZ4 \cite{Masui2015}}    & 2015           & HPC                       & {S,D}               & CPU             & SIMD + threads             & C+Python           & transform + dict. & open-source
	\\
	{bitshuffle::zstd \cite{Masui2015}}   & 2015           & HPC                       & {S,D}               & CPU             & SIMD + threads             & C+Python           & transform + dict. & open-source
	\\
	{Gorilla \cite{Pelkonen2015}}         & 2015           & Database                  & {\ \ D}             & CPU             & serial                   & go                 & delta                & open-source
	\\
	{SPDP \cite{Claggett2018}}            & 2018           & HPC                       & {S,D}               & CPU             & serial                   & C                  & dictionary           & open-source
	\\
	{ndzip-CPU \cite{Knorr2021}}          & 2021           & HPC                       & {S,D}               & CPU             & SIMD + threads             & C++                & transform+Lorenzo    & open-source
	\\
	{BUFF \cite{Liu2021}}                 & 2021           & Database                  & {S,D}               & CPU             & serial                   & rust               & delta                & open-source
	\\
	{Chimp\cite{liakos2022chimp}}         & 2022           & Database                  & {S,D}               & CPU             & serial                   & go                 & delta                & open-source
	\\
	\cline{2-9}
	{GFC \cite{o2011floating}}            & 2011           & HPC                       & {\ \ D}             & GPU             & SIMT                     & CUDA C             & delta                & open-source
	\\
	{MPC \cite{yang2015mpc}}              & 2015           & HPC                       & {S,D}               & GPU             & SIMT                     & CUDA C             & transform+delta      & open-source
	\\
	{nvcomp::LZ4 \cite{githubnvcomp}}     & 2020           & general                   & {S,D}               & GPU             & SIMT                     & CUDA C++           & transform + dict. & \itshape proprietary
	\\
	{nvcomp::bitcomp \cite{githubnvcomp}} & 2020           & general                   & {S,D}               & GPU             & SIMT                     & CUDA C++           & transform + prediction & \itshape proprietary
	\\
	{ndzip-GPU \cite{knorr2021ndzip}}     & 2021           & HPC                       & {S,D}               & GPU             & SIMT                     & SYCL C++           & transform + Lorenzo    & open-source
	\\
	{Dzip \cite{Goyal2021}}               & 2021           & general                   & {S,D}               & GPU             & SIMT                     & Pytorch            & prediction           & open-source
	\\
\cline{2-9}
\end{tabu}

%% file: tables/buffbits.tex
\small
\begin{tabular}{@{} |l *{10}{|c} | @{}}
    \hline
	\TabHead Precision   & 1 & 2 & 3  & 4  & 5  & 6  & 7  & 8  & 9  & 10 \\
    \hline
	\TabHead Bits needed & 5 & 8 & 11 & 15 & 18 & 21 & 25 & 28 & 31 & 35 \\
    \hline
\end{tabular}

%% file: tables/datasets_new.tex
\sffamily
\begin{tabu}{@{} r | >{\ttfamily}c | *{3}{>{\ttfamily}r | } @{}}
        \rowfont{\bfseries}
        \multicolumn{1}{r}{domain \& name}                          & \multicolumn{1}{c}{type$^*$}  & \multicolumn{1}{r}{size in bytes} & \multicolumn{1}{r}{entropy} & \multicolumn{1}{r}{extent}                                    \\
        \cline{2-5}
        \NameDomain{msg-bt}{HPC} \cite{sciendp}                     & {\ D} & 266,389,432   & 23.67   & 33298679 (1D)                             \\
        \NameDomain{num-brain}{HPC} \cite{sciendp}                  & {\ D} & 141,840,000   & 23.97   & 17730000 (1D)                             \\
        \NameDomain{num-control}{HPC} \cite{sciendp}                & {\ D} & 159,504,744   & 24.14   & 19938093 (1D)                             \\
        \NameDomain{rsim}{HPC} \cite{thoman2020rtx}                 & {S\ } & 94,281,728    & 18.50   & 2048$\,\times\,$11509 (2D)                \\
        \NameDomain{astro-mhd}{HPC} \cite{kissmann2016colliding}    & {\ D} & 548,458,560   & 0.97    & 130$\,\times\,$514$\,\times\,$1026 (3D)   \\
        \NameDomain{astro-pt}{HPC} \cite{huber2021relativistic}     & {\ D} & 671,088,640   & 26.32   & 512$\,\times\,$256$\,\times\,$640 (3D)    \\
        \NameDomain{miranda3d}{HPC} \cite{zhao2020sdrbench}         & {S\ } & 4,294,967,296 & 23.08   & 1024$\,\times\,$1024$\,\times\,$1024 (3D) \\
        \NameDomain{turbulence}{HPC} \cite{klacansky}               & {S\ } & 67,108,864    & 23.73   & 256$\,\times\,$256$\,\times\,$256 (3D)    \\
        \NameDomain{wave}{HPC} \cite{thoman2019celerity}            & {S\ } & 536,870,912   & 25.27   & 512$\,\times\,$512$\,\times\,$512 (3D)    \\
        \NameDomain{hurricane}{HPC} \cite{zhao2020sdrbench}         & {S\ } & 100,000,000   & 23.54   & 100$\,\times\,$500$\,\times\,$500 (3D)    \\
        \cline{2-5}
        \NameDomain{citytemp}{TS} \cite{citytemp}                  & {S\ } & 11,625,304    & 9.43    & 2906326 (1D)                              \\
        \NameDomain{ts-gas}{TS} \cite{fonollosa2015reservoir}      & {S\ } & 307,452,800   & 13.94   & 76863200 (1D)                             \\
        \NameDomain{phone-gyro}{TS} \cite{stisen2015smart}         & {\ D} & 334,383,168   & 14.77   & 13932632$\,\times\,$3 (2D)                \\
        \NameDomain{wesad-chest}{TS} \cite{schmidt2018introducing} & {\ D} & 272,339,200   & 13.85   & 4255300$\,\times\,$8 (2D)                 \\
        \NameDomain{jane-street}{TS} \cite{janestreet}             & {\ D} & 1,810,997,760 & 26.07   & 1664520$\,\times\,$136 (2D)               \\
        \NameDomain{nyc-taxi}{TS} \cite{yellowTaxi}                & {\ D} & 713,711,376   & 13.17   & 12744846$\,\times\,$7 (2D)                \\
        \NameDomain{gas-price}{TS} \cite{gasprice}                 & {\ D} & 886,619,664   & 8.66    & 36942486$\,\times\,$3 (2D)                \\
        \NameDomain{solar-wind}{TS} \cite{solarwind}               & {S\ } & 423,980,536   & 14.06   & 7571081$\,\times\,$14 (2D)                \\
        \cline{2-5}
        \NameDomain{acs-wht}{OBS} \cite{mastportal}                 & {S\ } & 225,000,000   & 20.13   & 7500$\,\times\,$7500 (2D)                 \\
        \NameDomain{hdr-night}{OBS} \cite{hdrphoto1}                & {S\ } & 536,870,912   & 9.03    & 8192$\,\times\,$16384 (2D)                \\
        \NameDomain{hdr-palermo}{OBS} \cite{hdrphoto2}              & {S\ } & 843,454,592   & 9.34    & 10268$\,\times\,$20536 (2D)               \\
        \NameDomain{hst-wfc3-uvis}{OBS} \cite{mastportal}           & {S\ } & 108,924,760   & 15.61   & 5329$\,\times\,$5110 (2D)                 \\
        \NameDomain{hst-wfc3-ir}{OBS} \cite{mastportal}             & {S\ } & 24,015,312    & 15.04   & 2484$\,\times\,$2417 (2D)                 \\
        \NameDomain{spitzer-irac}{OBS} \cite{spitzerirac}           & {S\ } & 164,989,536   & 20.54   & 6456$\,\times\,$6389 (2D)                 \\
        \NameDomain{g24-78-usb}{OBS} \cite{mastportal}              & {S\ } & 1,335,668,264 & 26.02   & 2426$\,\times\,$371$\,\times\,$371 (3D)   \\
        \NameDomain{jws-mirimage}{OBS} \cite{mastportal}            & {S\ } & 169,082,880   & 23.16   & 40$\,\times\,$1024$\,\times\,$1032 (3D)   \\
        \cline{2-5}
        \NameDomain{tpcH-order}{DB} \cite{tpcH}                    & {\ D} & 120,000,000   & 23.40   & 15000000 (1D)                             \\
        \NameDomain{tpcxBB-store}{DB} \cite{tpcxBB}                & {\ D} & 789,920,928   & 16.73   & 8228343$\,\times\,$12 (2D)                \\
        \NameDomain{tpcxBB-web}{DB} \cite{tpcxBB}                  & {\ D} & 986,782,680   & 17.64   & 8223189$\,\times\,$15 (2D)                \\
        \NameDomain{tpcH-lineitem}{DB} \cite{tpcH}                 & {S\ } & 959,776,816   & 8.87    & 59986051$\,\times\,$4 (2D)                \\
        \NameDomain{tpcDS-catalog}{DB} \cite{tpcDS}                & {S\ } & 172,803,480   & 17.34   & 2880058$\,\times\,$15 (2D)                \\
        \NameDomain{tpcDS-store}{DB} \cite{tpcDS}                  & {S\ } & 276,515,952   & 15.17   & 5760749$\,\times\,$12 (2D)                \\
        \NameDomain{tpcDS-web}{DB} \cite{tpcDS}                    & {S\ } & 86,354,820    & 17.33   & 1439247$\,\times\,$15 (2D)                \\
        \cline{2-5}
\end{tabu}

%% file: tables/comp_ratio_new.tex


\begin{tabu}{@{} r | rrrrrrrrr | rrrrr | @{}}
        \rowfont{\bfseries}
        \multicolumn{1}{r}{domain \& name} & 
        \TableHead{pFPC} & 
        \TableHead{SPDP} & 
        \TableHead{fpzip} & 
        \TableHead{\TwoLineTabTitle{shf+}{LZ4}} & 
        \TableHead{\TwoLineTabTitle{shf+}{zstd}} & 
        \TableHead{\TwoLineTabTitle{ndzip}{-CPU}} & 
        \TableHead{BUFF} & 
        \TableHead{Gorilla} & 
        \TableHead{Chimp} & 
        \TableHead{GFC} & 
        \TableHead{MPC} & 
        \TableHead{\TwoLineTabTitle{nv::}{LZ4}} & 
        \TableHead{\TwoLineTabTitle{nv::}{btcmp}} & 
        \TableHead{\TwoLineTabTitle{ndzip}{-GPU}} 
        \\
        \cline{2-15}
        \NameDomain{msg-bt}{HPC}     & 1.251     & 1.327   & 1.200   & 1.205   & 1.188   & 1.127  & 1.032 & 1.086  & 1.129 & 1.091 & 1.145 & 1.063  & 1.056  & 1.127  \\
        \NameDomain{num-brain}{HPC}     & 1.153     & 1.200   & 1.250   & 1.174   & 1.177   & 1.165  & 2.133 & 1.110  & 1.175 & 1.091 & 1.185 & 0.996  & 0.999  & 1.165  \\
        \NameDomain{num-control}{HPC}     & 1.036     & 1.011   & 1.120   & 1.114   & 1.117   & 1.109  & 2.207 & 0.980  & 1.057 & 1.013 & 1.108 & 1.013  & 1.009  & 1.109  \\
        \NameDomain{rsim}{HPC}     & 1.351     & 1.686   & 1.480   & 1.500   & 1.560   & 1.973  & 0.640 & 1.335  & 1.338 & 1.298 & 1.514 & 1.309  & 1.306  & 1.973  \\
        \NameDomain{astro-mhd}{HPC}     & 10.926    & 20.935  & 8.720   & 12.367  & 17.506  & 12.579 & 1.524 & 18.595 & 5.971 & \NA   & 8.132 & 22.824 & 20.801 & 12.579 \\
        \NameDomain{astro-pt}{HPC}     & 1.285     & 1.398   & 1.200   & 1.202   & 1.214   & 1.423  & 2.000 & 1.032  & 1.223 & \NA   & 1.224 & 0.996  & 0.999  & 1.423  \\
        \NameDomain{miranda3d}{HPC}     & 1.136     & 1.195   & 2.200   & \NA     & \NA     & 1.835  & 1.067 & 1.039  & 1.177 & \NA   & 1.495 & 1.020  & 1.019  & \NA    \\
        \NameDomain{turbulence}{HPC}     & 1.012     & 1.046   & 1.420   & 1.150   & 1.157   & 1.232  & 0.889 & 0.986  & 1.023 & 1.002 & 1.166 & 0.996  & 0.999  & 1.232  \\
        \NameDomain{wave}{HPC}     & 1.160     & 1.905   & 3.870   & 1.291   & 1.313   & 1.993  & 1.103 & 1.032  & 1.145 & 1.018 & 1.416 & 1.032  & 0.999  & 1.993  \\
        \NameDomain{hurricane}{HPC}     & 0.946     & 1.372   & \NA     & 1.513   & 1.552   & 0.974  & \NA   & 1.064  & 0.987 & 0.945 & 1.494 & 1.003  & 1.002  & 0.974  \\
        {\dm{\textbf{Domain-avg}}}& \dm{1.229}  & \dm{1.381} & \textbf{\dm{1.601}} & \dm{1.447}   & \dm{1.468}   & \dm{1.450}  & \dm{1.149} & \dm{1.161}  & \dm{1.232} & \dm{1.059} & \dm{1.399} & \dm{1.167}  & \dm{1.131}  & \textbf{\dm{1.420}}  \\
        \cline{2-15}
        \NameDomain{citytemp}{TS}      & 1.083     & 1.014   & 1.470   & 2.240   & 2.314   & 1.305  & 0.889 & 1.027  & 1.255 & 1.079 & 1.347 & 1.374  & 1.015  & 1.305  \\
        \NameDomain{ts-gas}{TS}      & 1.335     & 1.406   & 1.930   & 1.426   & 1.501   & 1.469  & 0.711 & 1.195  & 1.452 & 1.172 & 1.512 & 1.560  & 1.167  & 1.469  \\
        \NameDomain{phone-gyro}{TS}      & 1.031     & 1.083   & 1.060   & 1.199   & 1.243   & 1.000  & 1.939 & 0.971  & 1.384 & 1.023 & 1.190 & 1.808  & 0.999  & 1.000  \\
        \NameDomain{wesad-chest}{TS}      & 1.086     & 2.188   & 1.030   & 2.387   & 2.601   & 1.000  & 1.882 & 1.209  & 1.721 & 1.057 & 2.077 & 2.130  & 0.999  & 1.000  \\
        \NameDomain{jane-street}{TS}      & 1.034     & 1.000   & 1.080   & 1.066   & 1.032   & 1.087  & 1.600 & 0.968  & 1.025 & \NA   & 1.093 & 1.042  & 0.999  & 1.087  \\
        \NameDomain{nyc-taxi}{TS}      & 1.196     & 1.174   & 1.070   & 1.419   & 1.577   & 1.000  & 1.231 & 0.976  & 1.838 & \NA   & 1.098 & 1.836  & 1.004  & 1.000  \\
        \NameDomain{gas-price}{TS}      & 1.641     & 1.218   & 1.310   & 1.327   & 1.452   & 1.000  & 2.133 & 1.141  & 2.702 & \NA   & 1.204 & 2.895  & 0.999  & 1.000  \\
        \NameDomain{solar-wind}{TS}      & 0.956     & 1.108   & 1.040   & 1.116   & 1.113   & 1.000  & 0.627 & 0.968  & 1.083 & 0.968 & 1.051 & 1.172  & 0.999  & 1.000  \\
        {\dm{\textbf{Domain-avg}}}& \dm{1.148}  & \dm{1.235} & \dm{1.163} & \dm{1.334}   & \dm{1.387}   & \dm{1.061}  & \dm{1.176} & \dm{1.051}  & \textbf{\dm{1.457}} & \dm{1.050} & \dm{1.252} & \textbf{\dm{1.603}}  & \dm{1.020}  & \dm{1.061}  \\
        \cline{2-15}
        \NameDomain{acs-wht}{OBS}     & 1.220     & 1.252   & 1.640   & 1.468   & 1.488   & 1.478  & 0.727 & 1.251  & 1.226 & 1.231 & 1.491 & 1.165  & 1.165  & 1.478  \\
        \NameDomain{hdr-night}{OBS}     & 1.049     & 2.008   & 1.400   & 2.974   & 3.137   & 1.092  & 0.681 & 1.407  & 1.257 & 1.052 & 2.583 & 1.404  & 1.134  & 1.092  \\
        \NameDomain{hdr-lalermo}{OBS}     & 1.106     & 2.079   & 1.840   & 3.846   & 4.071   & 1.337  & 1.000 & 1.467  & 1.386 & \NA   & 3.713 & 1.418  & 1.359  & 1.337  \\
        \NameDomain{hst-wfc3-uvis}{OBS}     & 1.536     & 1.577   & 1.620   & 1.721   & 1.777   & 1.700  & 0.821 & 1.553  & 1.485 & 1.545 & 1.760 & 1.539  & 1.609  & 1.700  \\
        \NameDomain{hst-wfc3-ir}{OBS}     & 1.560     & 1.532   & 1.800   & 1.770   & 1.841   & 1.745  & 0.744 & 1.497  & 1.528 & 1.532 & 1.839 & 1.495  & 1.519  & 1.745  \\
        \NameDomain{spitzer-irac}{OBS}     & 1.186     & 1.229   & 1.320   & 1.359   & 1.379   & 1.304  & 0.821 & 1.196  & 1.178 & 1.191 & 1.346 & 1.234  & 1.235  & 1.304  \\
        \NameDomain{g24-78-usb}{OBS}     & 0.977     & 0.992   & 1.120   & 1.132   & 1.132   & 1.086  & 1.000 & 0.968  & 0.986 & \NA   & 1.103 & 0.996  & 0.999  & 1.086  \\
        \NameDomain{jws-mirimage}{OBS}     & 1.151     & 1.051   & 1.340   & 1.289   & 1.312   & 1.255  & 0.615 & 1.013  & 1.116 & 1.068 & 1.332 & 0.997  & 0.999  & 1.255  \\
        {\dm{\textbf{Domain-avg}}}& \dm{1.193}  & \dm{1.370} & \dm{1.471} & \dm{1.660}   & \textbf{\dm{1.697}}   & \dm{1.337}  & \dm{0.780} & \dm{1.258}  & \dm{1.246} & \dm{1.240} & \textbf{\dm{1.653}} & \dm{1.248}  & \dm{1.218}  & \dm{1.337}  \\
        \cline{2-15}
        \NameDomain{tpcH-order}{DB}      & 1.025     & 1.016   & 1.170   & 1.305   & 1.299   & 1.105  & 1.333 & 1.083  & 1.575 & 1.072 & 1.122 & 1.502  & 0.999  & 1.105  \\
        \NameDomain{tpcxBB-store}{DB}      & 1.084     & 1.095   & 1.080   & 1.477   & 1.537   & 1.000  & 1.488 & 0.980  & 2.227 & \NA   & 1.067 & 1.733  & 0.999  & 1.000  \\
        \NameDomain{tpcxBB-web}{DB}      & 1.081     & 1.098   & 1.090   & 1.458   & 1.477   & 1.000  & 1.455 & 0.982  & 2.169 & \NA   & 1.067 & 1.692  & 0.999  & 1.000  \\
        \NameDomain{tpcH-lineitem}{DB}      & 1.018     & 1.017   & 1.090   & 1.309   & 1.446   & 1.000  & 0.711 & 0.983  & 1.616 & \NA   & 1.010 & 1.510  & 0.999  & 1.000  \\
        \NameDomain{tpcDS-catalog}{DB}      & 0.982     & 0.998   & 1.090   & 1.106   & 1.117   & 1.000  & 0.727 & 0.970  & 1.027 & 0.976 & 1.034 & 1.058  & 0.999  & 1.000  \\
        \NameDomain{tpcDS-store}{DB}      & 0.988     & 0.990   & 1.070   & 1.096   & 1.129   & 1.000  & 0.744 & 0.973  & 1.049 & 0.990 & 1.033 & 1.134  & 0.999  & 1.000  \\
        \NameDomain{tpcDS-web}{DB}      & 0.987     & 0.998   & 1.100   & \NA     & \NA     & 1.000  & 0.727 & 0.970  & 1.026 & 0.976 & 1.034 & 1.057  & 0.999  & 1.000  \\
        {\dm{\textbf{Domain-avg}}}& \dm{1.022}  & \dm{1.029} & \dm{1.098} & \dm{1.274}   & \dm{1.313}   & \dm{1.014}  & \dm{0.920} & \dm{0.990}  & \textbf{\dm{1.382}} & \dm{1.002} & \dm{1.051} & \textbf{\dm{1.328}}  & \dm{0.999}  & \dm{1.014}  \\
        \cline{2-15}
        {Overall-avg} & 1.154     & 1.256   & 1.329 & 1.430 & \textbf{1.466} & 1.219  & 0.984 & 1.116  & 1.309 & 1.089 & \textbf{1.322} & 1.296  & 1.094  & 1.206  \\
        \cline{2-15}
\end{tabu}

%% file: tables/all_avg_throughputs.tex
\begin{tabular}{@{} r | rrrrrrrrr|rrrrr @{}}
\bfseries Metrics       & \bfseries pFPC    & \bfseries SPDP   & \bfseries fpzip  & \bfseries shf+LZ4 & \bfseries shf+zstd & \bfseries ndzip-C & \bfseries BUFF    & \bfseries Gorilla  & \bfseries Chimp & \bfseries GFC     & \bfseries MPC & \bfseries nv::LZ4 & \bfseries nv::btcomp  & \bfseries ndzip-G
	\\
	\hline
	\textbf{avg. comp}   & 0.564               & 0.181             & 0.079           & 0.923 & 1.407 & \textbf{2.192} & 0.202 & 0.047 & 0.034 & 87.778  & 29.595 & 2.716  & \textbf{240.280} & 142.635 \\
	\textbf{avg. decomp} & 0.351               & 0.178             & 0.074           & 1.181 & 1.328 & \textbf{1.636} & 0.254 & 0.146 & 0.175 & 99.258 & 28.513 & 53.352 & 122.483 & \textbf{159.312} \\
\end{tabular}

%% file: tables/end_to_end.tex
\begin{tabular}{@{} r | rrrrrrrrr|rrr @{}}
	\bfseries Metrics       & \bfseries pFPC    & \bfseries SPDP     & \bfseries fpzip
	                        & \bfseries shf+LZ4 & \bfseries shf+zstd & \bfseries ndzip-C
	                        & \bfseries BUFF    & \bfseries Gorilla  & \bfseries Chimp
	                        & \bfseries GFC     & \bfseries MPC
	                        & \bfseries ndzip-G
	\\
	\hline
	\textbf{avg. comp}   & 1602              & 2985               & 7103              & 403 & 328 & 282 & 2876 & 13760 & 16030 & \textbf{157} & 296 & 636 \\
	\textbf{avg. decomp} & 2104              & 2898               & 7368              & 365 & 347 & 334 & 2256 & 5498  & 3126  & \textbf{140} & 387 & 688 \\
\end{tabular}

%% file: tables/comp_ratio_md1d.tex
\begin{tabu}{@{} r *{5}{|c|c} | @{}}
        \rowfont{\bfseries}
        \multicolumn{1}{c}{}    & \multicolumn{2}{c}{GFC}     & \multicolumn{2}{c}{MPC}  & \multicolumn{2}{c}{fpzip} & \multicolumn{2}{c}{ndzip-C} & \multicolumn{2}{c}{ndzip-G}
        \\
        \cline{2-11}
            & md          & 1d          & md          & 1d          & md        & 1d & md & 1d & md & 1d
        \\
        \cline{2-11}
        \cline{2-11}
        \textbf{harmonic mean}                & 1.091                     & 1.089                     & 1.347                     & 1.365                     & 1.334                     & 1.326              & 1.223                & 1.210                & 1.207                & 1.200                \\
        \cline{2-11}
        \textbf{$p$-value    ($\alpha=0.05$)} & \multicolumn{2}{c|}{0.957} & \multicolumn{2}{c|}{0.691} & \multicolumn{2}{c|}{0.952} & \multicolumn{2}{c|}{0.848} & \multicolumn{2}{c|}{0.910}                                                                                                                  \\
        \cline{2-11}
\end{tabu}

%% file: tables/blocksizes.tex
\begin{tabu}{@{} ll rrrrrr|rr @{}}
    \bfseries blocksize & \bfseries metrics      & \bfseries pFPC
                        & \bfseries SPDP         & \bfseries shf+LZ4
                        & \bfseries shf+zstd
                        & \bfseries Gorilla
                        & \bfseries Chimp
                        & \bfseries nv::LZ4
                        & \bfseries nv::btcmp
    \\
    \hline
    \multirow{3}{*}{\textbf{4K}}
                        & \textbf{avg-CR}        & 1.151             & 1.215          & 1.426          & 1.463          & \textbf{1.116} & 1.309          & 1.244           & 1.075            \\
                        & \textbf{avg-CT (GB/s)} & 0.018             & 0.142          & 1.342          & 1.271          & 0.092          & 0.081          & 4.953           & 108.983          \\
                        & \textbf{avg-DT (GB/s)} & 0.017             & 0.143          & 1.179          & 1.190          & 0.192          & 0.226          & 88.669          & 71.286           \\
    \hline
    \multirow{3}{*}{\textbf{64K}}
                        & \textbf{avg-CR}        & \textbf{1.156}    & 1.239          & \textbf{1.430} & 1.466          & 1.095          & \textbf{1.315} & \textbf{1.296}  & \textbf{1.094}   \\
                        & \textbf{avg-CT (GB/s)} & 0.199             & \textbf{0.254} & \textbf{2.734} & \textbf{2.505} & 0.120          & 0.090          & 2.716           & \textbf{240.280} \\
                        & \textbf{avg-DT (GB/s)} & 0.129             & \textbf{0.250} & 2.849          & 3.463          & 0.332          & 0.260          & 53.352          & \textbf{122.483} \\
    \hline
    \multirow{3}{*}{\textbf{8M}}
                        & \textbf{avg-CR}        & 1.154             & \textbf{1.256} & 1.361          & \textbf{1.491} & 1.086          & \textbf{1.315} & 1.226           & 1.057            \\
                        & \textbf{avg-CT (GB/s)} & \textbf{0.640}    & 0.181          & 1.384          & 1.807          & \textbf{0.158} & \textbf{0.104} & \textbf{10.402} & 68.033           \\
                        & \textbf{avg-DT (GB/s)} & \textbf{0.405}    & 0.178          & \textbf{4.467} & \textbf{4.271} & \textbf{0.452} & \textbf{0.294} & \textbf{94.400} & 50.279           \\
    \hline
\end{tabu}

%% file: tables/query-from-h5.tex
\begin{tabu}{@{} r | rrrrrrrr|rrr|r |@{}}
\rowfont{\bfseries}
        \multicolumn{1}{r}{name} & pFPC    & SPDP     & fpzip
                                 & shf+LZ4 & shf+zstd & ndzip-C
                                 & Gorilla & \multicolumn{1}{r}{Chimp}
                                 & GFC     & MPC
                                 & \multicolumn{1}{r}{ndzip-G}
                                 & \multicolumn{1}{r}{query}
        \\
        \cline{2-13}
        {tpcH-order}      & 78+ 356           & 85+ 622            & 74+ 1577          & 66+ 94  & 66+ 72       & 74+ 60  & 75+1271  & 58+ 837  & 73+ 80  & 71+ 84       & 75+1232  & 190 \\
        {tpcxBB-store}    & 423+2032          & 491+3476           & 423+10162         & 311+657 & 299+623      & 463+712 & 472+5730 & 204+4531 & \NA     & 430+549      & 462+2024 & 268 \\
        {tpcxBB-web}      & 549+2522          & 612+4643           & 539+12845         & 387+777 & 378+778      & 597+892 & 612+8799 & 266+5748 & \NA     & 556+685      & 597+2224 & 292 \\
        {tpcH-lineitem}   & 565+2447          & 640+7463           & 525+14649         & 426+758 & 378+808      & 579+864 & 593+3102 & 339+6153 & \NA     & 576+669      & 576+2196 & 885 \\
        {tpcDS-catalog}   & 109+ 499          & 121+1145           & 100+ 2910         & 96+135  & 97+149       & 106+168 & 110+1329 & 105+1131 & 108+115 & 103+119      & 108+1372 & 64  \\
        {tpcDS-store}     & 161+ 757          & 188+1735           & 150+ 4686         & 147+217 & 142+204      & 161+260 & 162+1558 & 151+1769 & 160+185 & 156+191      & 158+1405 & 106 \\
        {tpcDS-web}       & 65+ 272           & 67+ 580            & 60+ 1457          & \NA     & \NA          & 60+ 98  & 64+ 676  & 61+ 568  & 63+ 58  & 62+ 68       & 58+1231  & 43  \\
        \cline{2-13}
        {arithmetic mean} & 1548              & 3162               & 7165              & 679     & {666} & 727     & 3508     & 3131     & 211     & {616} & 1960     & 264 \\
        \cline{2-13}
\end{tabu}